\documentclass[twoside,epsf,11pt]{article}
\usepackage{graphicx}
\usepackage{multicol}
\newcommand{\sW}{s_{\rm W}}

\font\tenrsfs=rsfs10
\font\sevenrsfs=rsfs7
\font\fiversfs=rsfs5
\newfam\rsfsfam
\textfont\rsfsfam=\tenrsfs
\scriptfont\rsfsfam=\sevenrsfs
\scriptscriptfont\rsfsfam=\fiversfs
\def\mathscr#1{{\fam\rsfsfam\relax#1}}
\newcommand{\Lag}{\mathscr{L}\,}

\newcommand{\NP}{Nucl. Phys.}

\newcommand{\PRL}{Phys. Rev. Lett.}
\newcommand{\PL}{Phys. Lett.}
\newcommand{\PR}{Phys. Rev.}
\def\SU{{\rm SU}}

\def\MeV{\,{\rm MeV}}
\def\circa#1{\,\raise.3ex\hbox{$#1$\kern-.75em\lower1ex\hbox{$\sim$}}\,}

\newcount\Mac  \Mac=1  
\newcommand{\ifMac}[2]{\ifnum\Mac=1 #1 \else #2 \fi}

\def\One{\hbox{1\kern-.24em I}}

\newcommand{\eV}{\,{\rm eV}}
\newcommand{\GeV}{\,{\rm GeV}}
\newcommand{\TeV}{\,{\rm TeV}}

\newcommand{\ds}{\partial\!\!\!\raisebox{2pt}[0pt][0pt]{$\scriptstyle/$}}

\newcommand{\eq}[1]{~{\rm (\ref{eq:#1})}}

\def\Red{}
\def\Black{}
\def\Blue{}

\newcounter{alphaequation}[equation]
\def\thealphaequation{\theequation\hbox to
0.6em{\hfil\alph{alphaequation}\hfil}}
\def\eqnsystem#1{
\def\@eqnnum{{\rm (\thealphaequation)}}
\def\@@eqncr{\let\@tempa\relax \ifcase\@eqcnt \def\@tempa{& & &} \or
  \def\@tempa{& &}\or \def\@tempa{&}\fi\@tempa
  \if@eqnsw\@eqnnum\refstepcounter{alphaequation}\fi
\global\@eqnswtrue\global\@eqcnt=0\cr}
\refstepcounter{equation} \let\@currentlabel\theequation \def\@tempb{#1}
\ifx\@tempb\empty\else\label{#1}\fi
\refstepcounter{alphaequation}
\let\@currentlabel\thealphaequation
\global\@eqnswtrue\global\@eqcnt=0 \tabskip\@centering\let\\=\@eqncr
$$\halign to \displaywidth\bgroup \@eqnsel\hskip\@centering
$\displaystyle\tabskip\z@{##}$&\global\@eqcnt\@ne
\hskip2\arraycolsep\hfil${##}$\hfil& \global\@eqcnt\tw@\hskip2\arraycolsep
$\displaystyle\tabskip\z@{##}$\hfil
\tabskip\@centering&\llap{##}\tabskip\z@\cr}
\def\endeqnsystem{\@@eqncr\egroup$$\global\@ignoretrue} \makeatother
\makeatletter
\def\art{\@ifnextchar[{\eart}{\oart}}
\def\eart[#1]#2#3#4#5#6{{\rm #2}, {\em #3 \bf #4} {\rm (#6) #5} ({\em #1})}
\def\hepart[#1]#2{{\rm #2, \em#1}}
\newcommand{\oart}[5]{{\rm #1}, {\em #2 \bf #3} {\rm (#5) #4}}

\def\beq{\begin{equation}}
\def\eeq{\end{equation}}
\def\bea{\begin{eqnarray}}
\def\eea{\end{eqnarray}}
\def\bq{\begin{quote}}
\def\eq{\end{quote}}

\def \lsim{\mathrel{\vcenter
     {\hbox{$<$}\nointerlineskip\hbox{$\sim$}}}}

\def\gappeq{\mathrel{\rlap {\raise.5ex\hbox{$>$}}
{\lower.5ex\hbox{$\sim$}}}}

\def\lappeq{\mathrel{\rlap{\raise.5ex\hbox{$<$}}
{\lower.5ex\hbox{$\sim$}}}}

\begin{document}

\centerline{  CERN--TH/2001--366\hfill
  IFUP--TH/2001-41\hfill
 IFIC/01-69 \hfill FTUV-01-1217 \hfill RM3-TH/2001-17}

\vspace{5mm}
\Black
\vspace{0.5cm}
\centerline{\huge\bf\Red Old and new physics interpretations }\vskip2mm
\centerline{\huge\bf\Red of the NuTeV anomaly}
\medskip\bigskip\Black
\medskip\bigskip\Black
 \centerline{\large\bf S. Davidson$^{a,b}$, S. Forte$^c$\footnote{On leave from INFN, Sezione di Torino, Italy}, P. Gambino$^d$,
N. Rius$^a$, A. Strumia$^{d\,2}$}\vspace{0.3cm}
\centerline{\em (a) Depto.\ de Fisica Te\'orica and IFIC, Universidad de Valencia-CSIC,
Valencia, Spain }
\centerline{\em (b) IPPP, University of Durham, Durham DH1 3LE,UK  }
\centerline{\em (c) INFN, Sezione di Roma III, Via della Vasca Navale,
I--00146, Roma, Italy}
\centerline{\em (d) Theoretical Physics Division, CERN, CH-1211 Gen\`eve 23, Suisse}
\vspace{0.4cm}

\vspace{1cm}
\Blue\centerline{\large\bf Abstract}
\begin{quote}\large\indent
We discuss whether the NuTeV anomaly can be explained,
compatibly with all other data,
by QCD effects (maybe, if the strange sea is asymmetric, or there is a
tiny violation of isospin),
new physics in propagators or couplings of the vector bosons (not really),
loops of supersymmetric particles (no),
dimension six operators (yes, for one specific $\SU(2)_L$-invariant operator),
leptoquarks (not in a minimal way),
extra U(1) gauge bosons (maybe: an unmixed $Z'$ coupled to $B-3L_\mu$
also increases the muon $g-2$ by about $10^{-9}$
and gives a `burst' to cosmic rays above the GZK cutoff).

\Black
\end{quote}
\vspace{5mm}

\footnotetext[2]{On leave from dipartimento di Fisica
dell'Universit\`a di Pisa and INFN.}

\section{Introduction}
The NuTeV collaboration~\cite{NuTeV} has recently reported a $\sim 3\sigma$ anomaly in
the NC/CC ratio of deep-inelastic $\nu_\mu$-nucleon scattering.
The effective $\nu_\mu$ coupling to left-handed quarks is found to be about $1\%$
lower than the best fit SM prediction.

As in the case of other apparent anomalies
(e.g. $\epsilon'/\epsilon$~\cite{eps'exp,trieste},
the muon $g-2$~\cite{g-2,lbl,had}, atomic parity
violation~\cite{Wood:1997zq,russiAPV}, and another puzzling NuTeV
result concerning dimuon events~\cite{dimuon},  to cite only the most recent cases)
one should first worry about theoretical uncertainties,
mainly due to QCD, before speculating on possible new physics.
After reviewing in section 2 the SM prediction for the NuTeV observables,
in section~3 we look for SM effects and/or uncertainties which could
alleviate the anomaly.
In particular, we investigate the possible effect of next-to-leading
order QCD corrections and consider the uncertainties related to
parton distribution functions (PDFs). We notice that
a small  asymmetry between strange and antistrange
in the quark  sea of the nucleon,
suggested by $\nu {\cal N}$ deep inelastic data~\cite{BPZ},
could be responsible for a significant fraction of the observed anomaly.
We  also study the effect  a very small violation of isospin symmetry
can have on the NuTeV result.

Having looked at the possible SM explanations, and keeping in mind that large
statistical fluctuations cannot be excluded, we then speculate
on the sort of physics beyond the SM that could be responsible for
the NuTeV anomaly. We  make a broad review of the main mechanisms
through which new physics may affect the quantities measured at NuTeV and
test them quantitatively, taking into account all the constraints
coming from other data.

We take the point of
view that interesting models should be able to explain a significant fraction
of the anomaly.
According to this criterion, we consider new physics that only affects the
propagators (section~\ref{oblique}) or gauge interactions
(section~\ref{couplings}) of the SM vector bosons, looking at the
constraints imposed on them by a global fit to the electroweak
precision observables.
Many models can generate a small fraction of the observed
discrepancy (see e.g.~\cite{Roy}), but it is more difficult to
explain a significant fraction of the anomaly.
In section~\ref{susy} we consider the case of the minimal
supersymmetric SM (MSSM) and look at possible MSSM quantum effects.
In section~\ref{NRO} we turn to  lepton-lepton-quark-quark
effective vertices,
focusing on the most generic set of dimension 6 operators.
We find that very few of them can fit the NuTeV anomaly
(in particular, only one $\SU(2)_L$-invariant operator).
In section~\ref{LQ} and~\ref{Z'} we study how these dimension six
operators could be generated by exchange of
leptoquarks or of extra U(1) gauge bosons.
Finally, we summarize  our findings in section 10.

\renewcommand{\arraystretch}{1.2}
\begin{table}
$$\begin{array}{|lccc|}\hline
\hbox{SM fermion}&{\rm U}(1)_Y&\SU(2)_L&\SU(3)_{\rm c}\cr \hline
U^c = u_R^c \phantom{*Ì{3\over 5}} & -{2 \over 3} & 1 & \bar{3} \cr
D^c = d_R^c \phantom{*Ì{3\over 5}} & \phantom{-}{1 \over 3}& 1 &\bar{3} \cr
E^c = e_R^c \phantom{*Ì{3\over 5}} &\phantom{-}1 & 1 &1  \cr
L=(\nu_L, e_L) & -{1 \over 2} & 2 &1\cr
Q=(u_L, d_L) &\phantom{-} {1\over 6}  & 2 & 3\cr \hline
\end{array}\qquad
\renewcommand{\arraystretch}{1.45}
\begin{array}{|c|cc|}\hline
Z\hbox{ couplings}
& g_L & g_R \\ \hline
\phantom{I}^{\phantom{I}^{\phantom{I}}}
\nu_e,\nu_\mu,\nu_\tau \phantom{I}^{\phantom{I}^{\phantom{I}}}
& \frac{1}{2} & 0 \\
\phantom{I}^{\phantom{I}^{\phantom{I}}}
e,\mu,\tau \phantom{I}^{\phantom{I}^{\phantom{I}}}
&-\frac{1}{2}+\sW^2 & \sW^2 \\
\phantom{I}^{\phantom{I}^{\phantom{I}}}
u,c,t \phantom{I}^{\phantom{I}^{\phantom{I}}}
& \phantom{-}\frac{1}{2}-\frac{2}{3}\sW^2 & -\frac{2}{3}\sW^2  \\
\phantom{I}^{\phantom{I}^{\phantom{I}}}
d,s,b \phantom{I}^{\phantom{I}^{\phantom{I}}}
& -\frac{1}{2}+\frac{1}{3}\sW^2 & \frac{1}{3}\sW^2 \\
\hline
\end{array}$$
\caption{\em The SM fermions and their $Z$ couplings.\label{tab:gAi}}
\end{table}
\renewcommand{\arraystretch}{1}

\section{The SM prediction}
\subsubsection*{Tree level}
In order to establish the notation and to present the physics in a simple approximation,
it is useful to recall the tree-level SM prediction for neutrino--nucleon deep inelastic scattering.
The $\nu_\mu$-quark effective Lagrangian predicted by the SM at tree level is
$$\Lag_{\rm eff} = -
2\sqrt{2}G_F ([\bar{\nu}_\mu \gamma_\alpha \mu_L ][\bar{d}_L \gamma^\alpha u_L ] +\hbox{h.c.}) -
2\sqrt{2}G_F \sum_{A,q} g_{Aq}
[\bar{\nu}_\mu\gamma_\alpha \nu_\mu][\bar{q}_A \gamma^\alpha q_A]
$$
where 
$A=\{L,R\}$, $q=\{u,d,s,\ldots\}$ and the $Z$ couplings
$g_{Aq}$ are given in table~\ref{tab:gAi} in terms of the
weak mixing angle $\sW\equiv\sin \theta_{\rm W}$.

It is convenient to define the ratios of neutral--current (NC) to
charged--current (CC)
deep-inelastic neutrino--nucleon
scattering total cross--sections $R_\nu$, $R_{\bar \nu}$.
Including only first generation quarks, for an
isoscalar target, and to leading order, these are given by
\begin{eqnarray}
R_\nu &\equiv& \frac{\sigma(\nu {\cal N}\to \nu X)}{\sigma(\nu {\cal N}\to \mu X)} =
\frac{(3 g_L^2 + g_R^2)q +  (3 g_R^2 + g_L^2)\bar q}{3 q +\bar q} = g_L^2 + r g_R^2\\
R_{\bar{\nu}} &\equiv& \frac{\sigma(\bar\nu {\cal N}\to \bar\nu X)}{\sigma(\bar\nu {\cal N}\to \bar\mu X)} =
 \frac{(3 g_R^2 + g_L^2)q +(3g_L^2 + g_R^2)\bar q}{q +3\bar q} = g_L^2 + \frac{1}{r} g_R^2,\label{rdef}
\end{eqnarray}
where $q$ and $\bar q$ denote  the second moments of quark or
antiquark distributions and correspond to the fraction of the nucleon
momentum carried by quarks and antiquarks, respectively.
For an isoscalar target, $q=(u+d)/2$, and
we have defined
\begin{equation}
r \equiv  \frac{\sigma(\bar{\nu}{\cal N}\to \bar\mu
X)}{\sigma({\nu}{\cal N}\to \mu X)}
=\frac{3 \bar{q} +q}{3q+\bar{q}}
\end{equation}
and
\begin{equation}
g_L^2 \equiv g_{Lu}^2 + g_{Ld}^2 = \frac{1}{2}-\sin^2\theta_{\rm W}+\frac{5}{9}\sin^4\theta_{\rm W},\qquad
g_R^2\equiv  g_{Ru}^2 + g_{Rd}^2 = \frac{5}{9}\sin^4\theta_{\rm W}.
\end{equation}

The observables $R_\nu^{\rm exp}$ and $R_{\bar{\nu}}^{\rm exp}$
measured at NuTeV differ from the expressions given in
eq.~(\ref{rdef}). On  the theoretical side this is due to
contributions from second--generation quarks, and
because of QCD and electroweak corrections. On the experimental
side, this is because total cross--sections can only be determined up
to  experimental cuts and uncertainties, such as those related to
the spectrum of the neutrino beam, the contamination of the $\nu_\mu$
beam by electron neutrinos,  and
the efficiency of NC/CC discrimination.
Once all these effects are  taken into account,
the NuTeV data can be viewed as a
measurement of the ratios between the CC 
and the NC 
squared neutrino effective couplings. The values quoted in~\cite{NuTeV} are
\begin{equation}\label{NuTeVgLgR}
g_L^2 = 0.3005\pm 0.0014\qquad\hbox{and}\qquad
g_R^2=0.0310\pm0.0011,
\end{equation}
where errors include both statistical and systematic uncertainties.

The difference of the effective couplings $g^2_L-g^2_R$
(`Paschos--Wolfenstein ratio'~\cite{PW}) is
subject to smaller theoretical and systematic uncertainties than
the individual couplings. Indeed, using eq.~(\ref{rdef}) we get
\begin{equation}\label{eq:PW}
R_{\rm PW} \equiv\frac{R_\nu - r R_{\bar{\nu}}}{1-r} =
\frac{\sigma(\nu {\cal N}\to \nu X)-\sigma(\bar\nu {\cal N}\to
\bar\nu X)}{\sigma(\nu {\cal N}\to \ell X) - \sigma(\bar{\nu}{\cal N}\to \bar{\ell}X)}=
 g_L^2- g_R^2 = \frac{1}{2}-\sin^2 \theta_{\rm W},
\end{equation}
which is seen to be independent of $q$ and $\bar{q}$,
and therefore of the information on the partonic structure of the nucleon.
Also, $R_{\rm PW}$ is
expected to be less sensitive to the various corrections discussed above.

\begin{figure}[t]
\parbox[t]{8cm}{$$
\includegraphics[width=70mm,height=70mm]{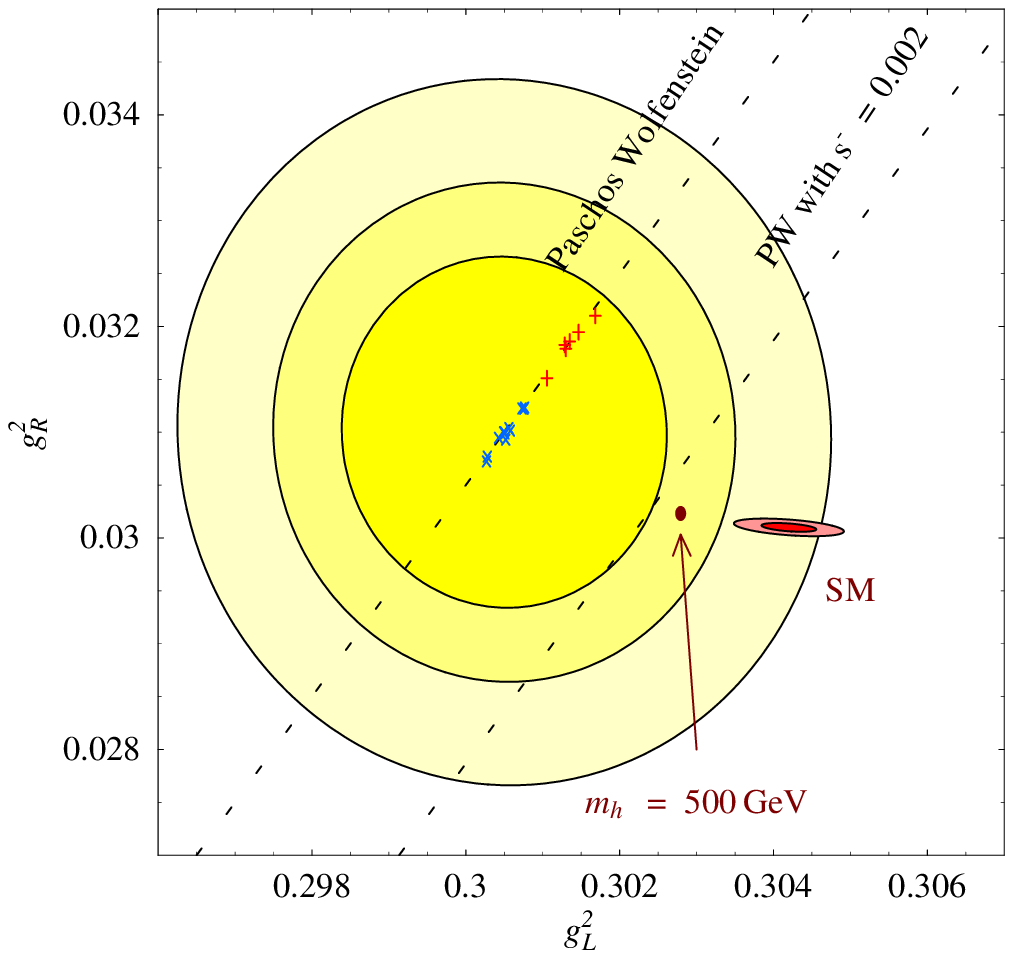} $$
\caption[]{\em  The SM prediction for ($g_L^2, g_R^2$) at $68,99\%$ CL and the
NuTeV determination, at $68,90,99\%$ CL.
The crosses show how the NuTeV central value moves
along the PW line using  different
sets of parton distribution functions that assume $s=\bar{s}$.
If $s>\bar s$  the PW line is shifted towards the SM
prediction.\label{plot1}}}\hfill
\parbox[t]{8cm}{$$
\includegraphics[width=70mm,height=70mm]{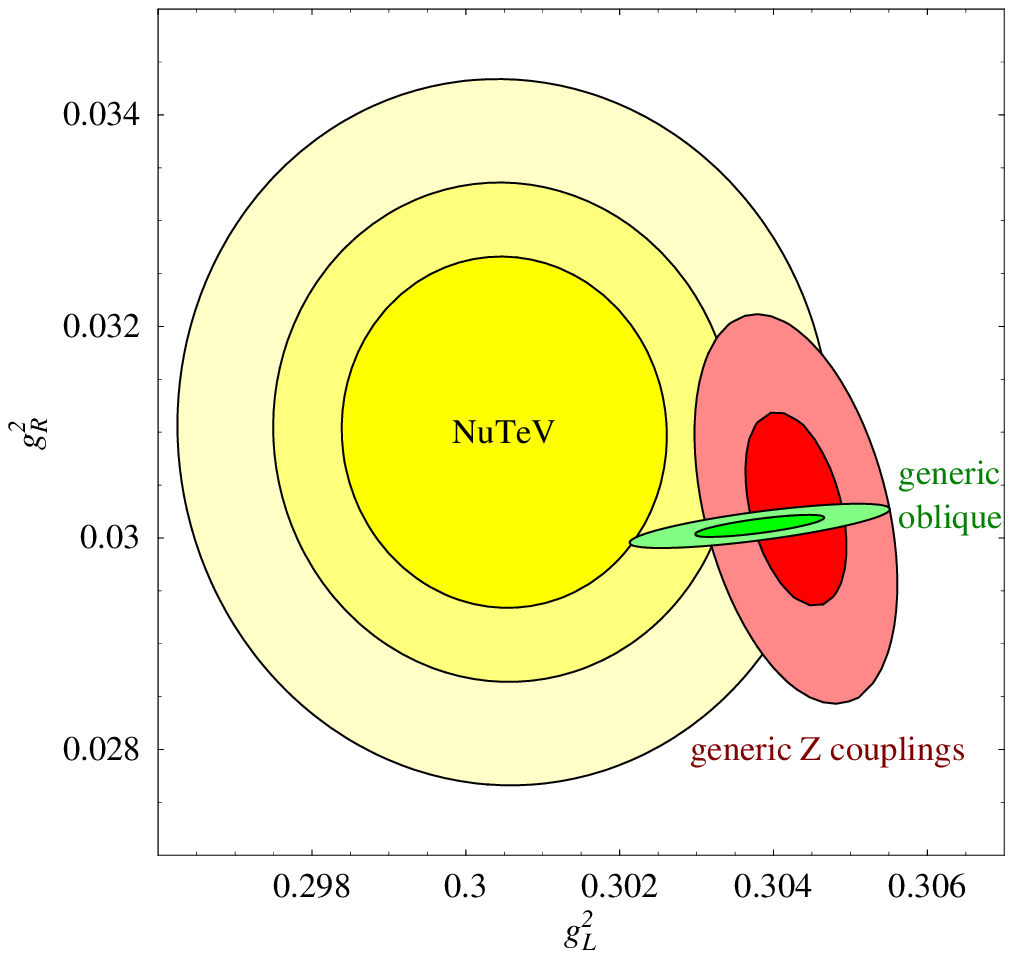} $$
\caption{\em
The NuTeV measurement is compared with the $68,99\%$ CL
ranges of ($g_L^2, g_R^2$) allowed by precision data
in classes of extensions of the SM that allow generic oblique corrections (almost horizontal
small green ellipses) or flavour-universal and $\SU(2)_L$-invariant corrections of the $Z$
couplings (vertical medium red ellipses).
\label{fig:plot2}}}
\end{figure}

\subsubsection*{Electroweak corrections and the SM fit}
The tree level SM predictions for $g_L$ and $g_R$
get  modified by electroweak radiative corrections.
These corrections depend on the precise definition of the weak mixing
angle, and we therefore adopt the {\em on-shell} definition~\cite{sirlin1980}
and define  $\sW^2\equiv 1-M_W^2/M_Z^2$.
One then obtains the following expressions for $g_{L,R}^2$~\cite{marciano80}
\begin{equation}
g_L^2= \rho^2 (\frac12- \sW^2 k + \frac59 \sW^4 k^2),\qquad
g_R^2=  \frac59 \rho^2  \sW^4 k^2,
\label{ewcorr}
\end{equation}
where, also including
the most important  QCD and electroweak higher order effects
\cite{paolo}
\begin{eqnarray*}
\rho&\approx& 1.0086 + 0.0001 (M_t/\GeV-175) -0.0006 \ln (m_h/100\GeV),\\
k&\approx& 1.0349 + 0.0004 (M_t/\GeV-175)-0.0029 \ln (m_h/100\GeV)
\end{eqnarray*}
for values of the top mass, $M_t$, and of Higgs mass,
$m_h$, not far from  $175$ and 100 GeV. Additional
very small non-factorizable terms are induced by electroweak box
diagrams.\footnote{QED radiative corrections depend sensitively
on the experimental setup and are taken into account in the NuTeV
analysis. The  charged current amplitude receives a small residual
renormalization which depends on the implementation of QED
corrections and is therefore neglected in our analysis.
}

We stress  that the SM value of $\sW$ depends on $M_t$ and $m_h$,
unless only the direct measurements of $M_W$ and $M_Z$ are used to compute it.
In particular, for fixed values of $M_t$ and $m_h$, the $W$ mass
can be very precisely determined from $G_F$, $\alpha(M_Z)$, and $M_Z$.
To very good approximation one  has (see e.g.\ \cite{degrassi})
\begin{equation}
M_W= 80.387 -0.058 \ln (m_h/100\GeV) -0.008 \ln^2 (m_h/100\GeV)
+ 0.0062 (M_t/\GeV-175).\nonumber
\end{equation}
Without including the NuTeV results,
the latest SM global fit of precision observables
gives $M_t=176.1\pm 4.3\GeV$,
$m_h=87^{+51}_{-34} \GeV$,
from which one obtains $M_W=80.400\pm0.019\GeV$~\cite{grunewald},
and therefore $\sW^2 = 0.2226\pm0.0004$.
The values of $g_L^2$ and $g_R^2$ corresponding to the best fit
are $0.3042$ and $0.0301$, respectively.
The small red ellipses in fig.~\ref{plot1}
show the SM predictions for $g_L^2$ and $g_R^2$ at $68\%$ and $99\%$ CL,
while the bigger yellow ellipses are the NuTeV data, at $68\%$ $90\%$ and $99\%$ CL.
While $g_R^2$ is in agreement with the
SM, $g_L^2$ shows a discrepancy of about $2.5\sigma$.

\smallskip

Here we have adopted the on-shell definition of $\sW$
because it is well known that with this choice
 the electroweak radiative corrections to $g_{L}^2$ cancel to a large
 extent.
In fact,  at first order in $\delta\rho\equiv \rho-1$
and $\delta k\equiv k-1$,
$g_L^2$ gets shifted by $\delta g_L^2\approx (2 \,\delta \rho- 0.551
\,\delta k) g_L^2$ and the leading quadratic dependence on $M_t$ is the
same for $ \delta \rho$ and $ \delta k\,\sW^2/c_{\rm W}^2 $. Therefore,
the top mass sensitivity of $g_L^2$ is very limited when
this effective coupling
 is expressed in terms of $\sW^2=1-M_W^2/M_Z^2$. As leading
higher order electroweak corrections are usually related to the high
value of the top mass~\cite{paolo},
higher order corrections cannot  have any relevant
impact on the discrepancy between the SM and NuTeV.

Within the SM, one can extract a value of $\sW^2$
from  the NuTeV data. This is performed by the NuTeV collaboration
using a fit to $\sW^2$ and the effective charm mass that is used to
describe  the
charm threshold. This fit is different from
the one that gives $g_{L,R}^2$. The result is $m_c^{\rm eff}=1.32\pm 0.11$
GeV and
\begin{eqnarray}\label{eq:stwres}
\sW^2&=&0.2276\pm0.0013~{\rm (stat.)}\pm0.0006~{\rm
  (syst.)}\pm0.0006~{\rm (th.)}\\
&& - 0.00003 (M_t/\GeV-175) + 0.00032 \ln (m_h/100\GeV) .
\nonumber
\end{eqnarray}
Here the systematic uncertainty  includes all the sources of
experimental systematics, such as those mentioned above, and it is
estimated by means of a Monte Carlo simulation of the experiment.
The theoretical
uncertainty is almost entirely given by QCD corrections, to be
discussed in section~3.
The total uncertainty above
is about 3/4 of that in the preliminary result from NuTeV
$\sW^2=0.2255 \pm 0.0021$~\cite{nutold} and about 1/2 of that in the
CCFR result $\sW^2=0.2236 \pm 0.0035$~\cite{CCFR}.

Alternatively, by equating eq.~(\ref{ewcorr}) to the NuTeV results for
$g_{L,R}^2$, we find
\begin{eqnarray}\label{eq:sWnuTeV}
\sW^2\hbox{(NuTeV})=
0.2272 \pm 0.0017 \pm 0.0001 \hbox{ (top)} \pm 0.0002
\hbox{ (Higgs)}.
\end{eqnarray}
The central value and the errors have been computed using
the best global fit for $M_t$ and $m_h$. Eq.~(\ref{eq:sWnuTeV})
has a slightly larger error but  is very close to
eq.~(\ref{eq:stwres}).
An additional difference between the two determinations
is that eq.~(\ref{eq:sWnuTeV})
is based on  a up-to-date treatment of higher order effects.
Notice that the NuTeV error is much larger than in the global fit given above,
from which eq.~(\ref{eq:stwres}) differs by about 3$\sigma$
and eq.~(\ref{eq:sWnuTeV}) by about $2.6\sigma$.
The NuTeV result for   $\sW^2$
can also be re-expressed in terms of  $M_W$. 
If we then compare with  $M_W=80.451\pm 0.033\GeV$ from direct measurements
at LEP and Tevatron, the discrepancy is even higher: more than
$3\sigma$ in both cases.
The inclusion of the NuTeV data in a global fit  shifts
the preferred $m_h$ value very slightly, but  worsens the
fit significantly ($\chi^2=30$ for 14 degrees of freedom).

Even without including NuTeV data, the global SM fit has a somewhat low
goodness-of-fit,  8\% if na\"{\i}vely estimated with a global Pearson
$\chi^2$ test.
The quality of the fit becomes considerably worse if only
the most precise data are retained \cite{altarelli}. Indeed,
 among the most precise observables,
the leptonic asymmetries measured at LEP and SLD and   $M_W$ point
to an extremely light Higgs, well below the direct exclusion bound
$m_h> 115 \GeV$,
while  the forward-backward hadronic asymmetries measured at LEP
prefer a very heavy Higgs (for a detailed discussion,
see~\cite{altarelli,chanowitz}). The effective leptonic couplings measured
by the hadronic asymmetries differ by more then 3$\sigma$ from those
measured by purely leptonic asymmetries.
Therefore,
the discrepancy between NuTeV and the other data depends also on
how this  other discrepancy  is treated.
For instance, a fit which excludes the hadronic asymmetries has
a satisfactory goodness-of-fit, but $m_h=40\GeV$ as best fit value.
In this case, the SM central values for
$g_{L,R}^2$ are 0.3046 and 0.0299, and differ even more from
the NuTeV measurements. On the other hand, even a very heavy Higgs
would not resolve the anomaly: to explain completely the
NuTeV result $m_h$ should be as heavy as 3~TeV, deep in the
non-perturbative regime. The preference of the NuTeV result for
a heavy Higgs is illustrated in fig.~\ref{plot1}
where we display the point corresponding to the SM predictions with
$m_h=500\GeV$ and $m_t=175\GeV$.
This is suggestive that,  as will be more clearly seen
in the following, the central value of NuTeV cannot be explained by radiative
corrections.

\section{QCD corrections}

Most of the  quoted
theoretical error on the NuTeV determination of $\sW^2$  is
due to QCD effects.
Yet, this uncertainty does not include some of the
assumptions on which the Paschos--Wolfenstein
relation, eq.~(\ref{eq:PW}), is based. Hence, one may ask: first,
whether some
source of violation of the Paschos--Wolfenstein relation which has not
been included in the experimental analysis can explain the observed
discrepancy, and second, whether some of the theoretical uncertainties
might actually be larger than estimated in~\cite{NuTeV}.

A full  next--to--leading order (NLO) treatment of neutrino deep--inelastic
 scattering is possible, since all the relevant coefficient functions
are long known~\cite{furpet}.  If no assumption on the parton content of
the target is made, including NLO corrections,
the Paschos--Wolfenstein ratio eq.~(\ref{eq:PW})
becomes
\begin{eqnarray}
R_{\rm PW}&=&g_L^2- g_R^2 \nonumber +{(u^--d^-)+(c^--s^-)\over{\cal Q}^-}
\Bigg\{\left[\frac{3}{2}(g_{Lu}^2-g_{Ru}^2) +
\frac{1}{2}(g_{Ld}^2-g_{Rd}^2)\right]+\\
&&+\frac{\alpha_s}{2\pi}(g_L^2-
g_R^2)(\frac{1}{4}\delta C^1-\delta C^3) \Bigg\}
+O({\cal Q}^-)^{-2}\label{eq:PWNLO}
\end{eqnarray}
The various quantities which enter eq.~(\ref{eq:PWNLO}) are defined as
follows:  $\alpha_s$ is the strong coupling; $\delta C^1\equiv C^1-C^2$,
$\delta C^3\equiv C^3-C^2$;   $C^i$ is the
the second moment of the  next--to--leading contributions to the
quark coefficient functions for structure function $F^i$;
$ q^-\equiv q-\bar q$;  ${\cal Q}^-\equiv (u^-+d^-)/{2}$;
$u$, $d$, and
so on are second moments of the corresponding quark and antiquark
distributions. We have expanded the result in powers of ${1/
{\cal Q}^-}$, since we are interested in the case of targets where the
dominant parton is the isoscalar ${\cal Q}^-=(u^-+d^-)/2$.
Equation~(\ref{eq:PWNLO}) shows the well-known fact
that the Paschos-Wolfenstein relation
is corrected if either the target has an isotriplet component
(i.e. $u\not=d$) or
sea quark contributions have a
$C$-odd component (i.e. $s^-\not=0$ or $c^-\not=0$).
Furthermore,  NLO corrections only affect these
isotriplet or $C$-odd terms.

Let us now consider these corrections in turn.
Momentum fractions are scale dependent; in the energy range of the
NuTeV experiment ${\cal Q}^-\approx 0.18$~\cite{MRST}, with better than
10\% accuracy, so that $\left[\frac{3}{2}(g_{Lu}^2-g_{Ru}^2) +
\frac{1}{2}(g_{Ld}^2-g_{Rd}^2)\right]/{\cal Q}^-\approx 1.3$. Hence, a
value $(u^--d^-)+(c^--s^-)\approx-0.0038$ is required to shift
the value of $\sW^2$ by an amount equal to the difference
between the NuTeV value central eq.~(\ref{eq:stwres}) and the global SM fit.

The NuTeV experiment uses an iron target, which has an excess of
neutrons over protons of about 6\%. This violation of isoscalarity is
however known to good accuracy, it is included in the NuTeV
analysis, and it gives a negligible contribution to the overall error.
A further violation of isoscalarity
could be due to the fact that isospin symmetry is violated by the
parton distributions of the nucleon, i.e. $u^p\not=d^n$ and
$u^n\not=d^p$. This effect is considered by
NuTeV~\cite{NuTeV}, but not included in their analysis.
Indeed, isospin in QCD is only violated by terms of order
$(m_u-m_d)/\Lambda$, and thus isospin violation  is
expected to be smaller than 1\% or so at the NuTeV scale
(where the scale dependence is rather weak)~\cite{ivln}.
However, if one were to conservatively
estimate the associated uncertainty by assuming isospin violation of
the valence distribution to be
at most  1\% (i.e. $(u^--d^-)/{\cal Q}^-\leq0.01$),
this would lead to a theoretical uncertainty on $\sW^2$ of
order $\Delta {\sW^2}=0.002$. This is  a more than threefold increase
in theoretical uncertainty, which would rather reduce the
significance of the NuTeV anomaly.

A $C$-odd second moment of heavier flavours,  $s^-\not=0$ or $c^-\not=0$ is
not forbidden by any symmetry of QCD, which only imposes that the
first moments of all heavy flavours must vanish. Neither of these
effects has been considered by NuTeV.
A nonzero value of
$c^-$ appears very unlikely since the perturbatively
generated charm component has $c^-=0$ identically for all moments,
and even assuming
that there is an `intrinsic' charm component (i.e.\ $c\not=0$ below
charm threshold due to nonperturbative effects) it is expected to have
vanishing $c^-$~\cite{intc} for all moments.
On the contrary, because the relevant
threshold  is
in the nonperturbative region, the strange
component is determined  by infrared dynamics and there is
no reason why $s^-=0$. In fact, explicit model calculations~\cite{ints} suggest
$s^-\approx 0.002$. Whereas such a $C$-odd strange component was
at first ruled out by CCFR dimuon data~\cite{ccfrs}\footnote{Deep-inelastic charm production
events are know as dimuon events because their experimental signature
is a pair of opposite--sign muons. Since they mainly proceed through
scattering of the neutrino off a strange quark, they are a
sensitive probe of the strange distribution.}, a subsequent
global fit to all available neutrino data found evidence in favor of
a strange component of this magnitude and sign~\cite{BPZ}, and showed
that it does not necessarily contradict the direct CCFR measurement.
More recent measurements~\cite{Goncharov} confirm the CCFR results in
a wider kinematic region; however, the quantitative impact of these data
on a global fit is unclear.
Even though it is not included in current parton sets, a
small asymmetry $s^-\approx 0.002$ seems
compatible with all the present experimental information~\cite{RGRp}.
Assuming $s^-\approx 0.002$ as suggested by~\cite{BPZ},
the value of $\sW^2$ measured by NuTeV
is lowered by about $\delta \sW^2 =0.0026$.
The corresponding shift of the PW line is displayed in fig.~\ref{plot1}.
This
reduces the discrepancy between NuTeV and the SM
to the level of about  one and a half standard deviations (taking
the NuTeV error at face value), thus eliminating the anomaly.

\smallskip

Since NLO corrections in eq.~(\ref{eq:PWNLO}) only affect the $C$-odd or
isospin-odd terms, they are in practice a sub--subleading
effect. Numerically, $\delta C^1- 4\,\delta C^3=16/9$ so NLO effects will
merely correct a possible isotriplet or $C$-odd contribution by making it
larger by a few percent.
Therefore, a purely leading-order analysis of $R_{\rm PW}$
is entirely adequate, and neglect of
NLO corrections should not contribute significantly to either the
central value of $\sW^2$ extracted from $R_{\rm PW}$, nor to the
error on it. It is important to realize, however, that this is not the
case when considering the individual ratios $R_\nu$ and
$R_{\bar\nu}$. Indeed, NLO corrections affect the leading--order
expressions by terms proportional to the dominant quark component
${\cal Q}^-$, and also by terms proportional to the gluon distribution,
which carries about 50\% of the nucleon's momentum.
 Therefore, one expects
NLO corrections to $R_\nu$ and $R_{\bar
\nu}$ to be of the same size as NLO corrections to typical observables
at this scale, i.e. around 10\%. The impact of this on the values of
$g^2_L$ and $g^2_R$, however, is difficult to assess: the NuTeV
analysis makes use of a parton set which has been self--consistently
determined fitting leading--order expressions to neutrino data, so
part of the NLO correction is in effect included in the parton
distributions. A reliable determination of $g^2_L$ and $g^2_R$ could
only be obtained if the whole NuTeV analysis were consistently
upgraded to NLO. As things stand, one should be aware that the NuTeV
determination of $g^2_L$ and $g^2_R$,  eq.~(\ref{NuTeVgLgR}), is
affected by a theoretical uncertainty  related to NLO which has not
been taken into account and which may well be non-negligible. This
uncertainty is however correlated between $g^2_L$ and $g^2_R$, and it
cancels when evaluating the difference $g^2_L-g^2_R$.

On top of explicit violations of the Paschos--Wolfenstein relation,
other sources of uncertainty are due to the fact that the experiment
of course does not quite measure  total cross--sections. Therefore,
some of the dependence on the structure of the nucleon which cancels
in ideal observables such as $R_\nu$ or $R_{\rm PW}$ remains in actual
experimental observables. In order to estimate these uncertainties,
we have developed a simple Monte Carlo which
simulates the  NuTeV experimental set-up. The Monte Carlo calculates
integrated cross sections with cuts typical of a $\nu{\cal N}$ experiment,
by using leading--order expressions.
Because the Monte Carlo is
not fitted self--consistently to the experimental raw data,
it is unlikely to give an accurate description of actual data.
However, it can be used to assess the uncertainties involved in
various aspects of the analysis.

We have therefore studied the variation of the result for $R_{\rm PW}$ as
several theoretical assumptions are varied, none of which affects the
ideal observable~$R_{\rm PW}$ but all of which affect the experimental
results. First, we have considered the dependence on parton
distributions. Although the error on parton distributions cannot
really be assessed at present, it is unlikely to be much larger than the
difference between leading--order and NLO parton sets. We can study
this variation by comparing the CTEQL and CTEQM parton
sets~\cite{cteq5}. We also compare results to those of the MRST99
set~\cite{MRST}. We find extremely  small variation for
$R_{\rm PW}$ and small variations even for the extraction of $g_{L,R}^2$.
Specific uncertainties which may affect significantly neutrino cross sections
are the relative size of up and down distributions at large
$x$~\cite{udrat,Kuhlmann:1999sf}
and the size of the strange and
charm component~\cite{chsiz}.
Both have been explored by MRST~\cite{MRST}, which produce parton sets
where all these features are varied in turn.
Again, using these parton sets in turn,
we find no significant variation of the
predicted $R_\nu$, $R_{\bar \nu}$, and of the extracted $g_{L,R}^2$.
If, on the contrary, we relax the assumption $s=\bar s$, which is
implicit in all these parton sets, we find a shift of $R_{\rm PW}$ in
good agreement with eq.~(\ref{eq:PWNLO}).
This conclusion
appears to be robust, and only weakly affected by the choice of parton
distributions  and by the specific $x$-dependence of the $s-\bar{s}$
difference, provided the second moment of $s-\bar{s}$ is kept fixed.
The lower cut ($20 \GeV$)
imposed by NuTeV on the energy deposited in the calorimeter tends to
decrease the sensitivity to the asymmetry $s^-$, as it mostly
eliminates high $x$ events. However, this effect is relevant only for
lower energy neutrinos, below about 100 GeV, and should be small in
the case of NuTeV.

The dependence on the choice of parton distributions
is shown in fig.~\ref{plot1} where blue $\times$ (red $+$) crosses correspond
to MRST99 (CTEQ) points. We cannot show a NuTeV value, because we
could not access
the parton set used by NuTeV.
The results are seen to spread  along the expected PW
line. The  intercept of this line
turns out to be determined by the input value of $g_L^2-g_R^2$, and to
be completely insensitive to details of parton distributions.
However,  it should be kept in mind that inclusion of NLO corrections
might alter significantly these results, by increasing the spread
especially
in the direction along the PW line, for the reasons discussed above.

Finally, we have tried to vary the charm mass, and to
 switch on some higher twist effects
(specifically those related to the nucleon mass). In both cases the
contributions to the uncertainty which we find are in agreement with
those of NuTeV.

\section{Oblique corrections}\label{oblique}
After our review of the SM analysis, let us proceed with a discussion
of possible effects of physics beyond the SM. We first concentrate on
new physics  which is characterized by a high mass scale and couples
only or predominantly to the vector bosons.
In this case  its contributions can be parameterized in
a model independent way by three ({\it oblique}) parameters.
Among the several equivalent parameterizations \cite{others},
we adopt $\epsilon_1,\epsilon_2,\epsilon_3$ \cite{eps}.
Many models of physics beyond the SM can be studied at
least approximately in this simple way.

Generic contributions to $\epsilon_1,\epsilon_2,\epsilon_3$ shift
$g_{L,R}^2$ according to the approximate expressions
\begin{equation}
\frac{\delta g_L^2}{ g_L^2}= 2.8 \,\delta \epsilon_1 -1.1 \,\delta
\epsilon_3;
\ \ \ \ \ \
\frac{\delta g_R^2}{ g_R^2}= -0.9 \,\delta \epsilon_1 +3.7 \,\delta
\epsilon_3.
\label{eps}
\end{equation}
Of course, the $\epsilon_i$ parameters are strongly constrained by electroweak
precision tests.
In order to see if this generic class of new physics
can give rise to the NuTeV anomaly, we extract the $\epsilon_i$
parameters directly from a fit to the electroweak data,
without using the SM predictions for them.
We use the most recent set of electroweak observables,
summarized in table~\ref{tab:data},
properly taking into account the uncertainties on $\alpha_{\rm
  em}(M_Z)$ and $\alpha_{\rm s}(M_Z)$. The result is a fit to
the $\epsilon_i$ very close to the one reported in~\cite{altarelli} which
we use in eq.~(\ref{eps}) after normalizing to the SM prediction at a
reference value of $m_h$.
The ellipses corresponding to the $68\%$ and $99\%$ CL are displayed
in fig.~\ref{fig:plot2} (green, almost horizontal ellipses).
They are centred roughly around the SM best fit point, because
the SM predictions for $\epsilon_1$ and $\epsilon_3$ for $m_h\approx 100$ GeV
are in reasonable agreement with the data (see also section~2).
The difference between the best fit point and the light Higgs SM
prediction for $(g_L^2,g_R^2)$ is much smaller than the NuTeV
accuracy. Notice that, as mentioned in section 2, excluding the hadronic
asymmetries from the fit would make an  oblique explanation even harder.

Our conclusion is that  oblique corrections cannot account for
the NuTeV anomaly, as they  can only absorb about $\sim1\sigma$ of the
 $\sim3\sigma$ NuTeV anomaly.

\begin{table}[t]
$$\begin{array}{rclrl}
G_{\rm F} &=& 1.16637~10^{-5}/\GeV^2&                      & \hbox{Fermi constant for $\mu$ decay}\\
M_Z &=& 91.1875 \GeV&                                       &\hbox{pole $Z$ mass}  \\
m_t  &=& (174.3\pm5.1)\GeV &                               &\hbox{pole top masss}\\
\alpha_{\rm s}(M_Z)  &=& 0.119\pm0.003 &                         &\hbox{strong coupling}\\
\alpha_{\rm em}^{-1}(M_Z)  &=& 128.936\pm0.046 &            & \hbox{electromagnetic coupling}\\
M_W &=& (80.451 \pm 0.033)\GeV        &  1.8\hbox{-}\sigma & \hbox{pole $W$ mass}  \\
\Gamma_Z &=& (2.4952 \pm 0.0023)\GeV  & -0.4\hbox{-}\sigma & \hbox{total $Z$ width} \\
\sigma_h &=&(41.540 \pm 0.037)\hbox{nb}&  1.6\hbox{-}\sigma & \hbox{$e\bar{e}$ hadronic cross section at $Z$ peak}\\
R_\ell &=& 20.767 \pm 0.025           &  1.1\hbox{-}\sigma & \hbox{$\Gamma(Z\to \hbox{hadrons})/\Gamma(Z\to\mu^+\mu^-)$}\\
R_b &=& 0.21646 \pm 0.00065           &  1.1\hbox{-}\sigma & \hbox{$\Gamma(Z\to b \bar b)/\Gamma(Z\to \hbox{hadrons})$}\\
R_c &=& 0.1719 \pm 0.0031             & -0.1\hbox{-}\sigma & \hbox{$\Gamma(Z\to c \bar c)/\Gamma(Z\to \hbox{hadrons})$}\\
A_{LR}^e &=& 0.1513 \pm 0.0021        &  1.6\hbox{-}\sigma & \hbox{Left/Right asymmetry in $e\bar{e}$}\\
A_{LR}^b &=& 0.922 \pm 0.02           & -0.6\hbox{-}\sigma &
\hbox{LR Forward/Backward asymmetry in $e\bar{e}\to b\bar{b}$}\\
A_{LR}^c &=& 0.670 \pm 0.026           &  0.1\hbox{-}\sigma & \hbox{LR
 FB asymmetry in $e\bar{e}\to c\bar{c}$}\\
A_{FB}^\ell &=& 0.01714 \pm 0.00095    &  0.8\hbox{-}\sigma & \hbox{Forward/Backward asymmetry in $e\bar{e}\to \ell\bar{\ell}$}\\
A_{FB}^b &=& 0.099 \pm 0.0017         & -2.8\hbox{-}\sigma & \hbox{Forward/Backward asymmetry in $e\bar{e}\to b\bar{b}$}\\
A_{FB}^c &=& 0.0685 \pm 0.0034        & -1.7\hbox{-}\sigma & \hbox{Forward/Backward asymmetry in $e\bar{e}\to c\bar{c}$}\\
Q_W &=& -72.5 \pm 0.7                 &  0.6\hbox{-}\sigma & \hbox{atomic parity violation in Cs}\\
\end{array}$$
\caption{\em The electroweak data included in our fit~\cite{LEPEWWG}.
The second column indicates the discrepancy with respect to the best SM fit.\label{tab:data}}
\end{table}

\section{Corrections to gauge boson interactions}\label{couplings}
We now discuss whether 
the NuTeV anomaly could be explained by modifying the
couplings of the vector bosons.
This possibility could work if new physics only affects
the $\bar{\nu}Z\nu$ couplings,
reducing the squared $\bar{\nu}_\mu Z \nu_\mu$ coupling by $(1.16\pm 0.42)\%$~\cite{NuTeV}.
This shift is consistent with precision
LEP data, that could not measure the $\bar{\nu}Z \nu$ couplings
as accurately as other couplings
(no knowledge of the LEP luminosity is needed to test charged lepton and quark couplings), and found
a $Z\to \nu\bar{\nu}$
rate $(0.53\pm 0.28)\%$ lower than the best-fit SM prediction.
We could not construct a  model
that naturally realizes this intriguing possibility, because
precision data test the $\bar{\mu} Z \mu$ and $\bar{\mu} W \nu_\mu$ couplings
at the per mille accuracy.
This generic problem is best understood by considering
explicit examples. We first show that models
where neutrinos mix with some extra fermion
(thereby shifting not only the $\bar{\nu} Z \nu$ coupling,
but also the $\bar{\ell} W \nu$ coupling)
do not explain the NuTeV anomaly.
Next, we discuss why a model where the $Z$ mixes with some extra vector boson
(thereby shifting not only the $\bar{\nu} Z \nu$ coupling, but also
the $Z$ couplings of other fermions)
does not explain the NuTeV anomaly.

\subsubsection*{Models that only affect the neutrino couplings}
This happens e.g.\ in models where the SM neutrinos
mix with right-handed neutrinos
(a $1\%$ mixing could be naturally obtained in extra dimensional models
or in appropriate 4-dimensional models~\cite{NuMixing}).
By integrating out the right-handed neutrinos,
at tree level one obtains the low energy effective lagrangian
\begin{equation}\label{eq:nud}
\Lag_{\rm eff} = \Lag_{\rm SM} +\epsilon_{ij}~
2\sqrt{2}G_F (H^\dagger \bar{L}_i)i\ds (HL_j),
\end{equation}
where $i,j=\{e,\mu,\tau\}$,
$L_i$ are the lepton left-handed doublets, $H$ is the Higgs doublet,
and $\epsilon_{ij}=\epsilon_{ji}^*$ are dimensionless couplings.
This peculiar dimension~6 operator only affects neutrinos.
After electroweak symmetry breaking, it
affects the kinetic term of the  neutrinos,
that can be recast in the standard form with
a redefinition of the neutrino field.
In this way, the $\bar{\nu}_i Z \nu_i$ and the
$\bar{\nu}_i W \ell_i$
couplings become respectively
$1-\epsilon_{ii}$ and $1-\epsilon_{ii}/2$
lower than in the SM ($\epsilon_{ii}$ is positive:
gauge couplings of neutrinos get reduced if neutrinos mix with neutral singlets).
The NuTeV anomaly would require $\epsilon_{\mu\mu}=0.0116\pm 0.0042$.
However, a reduction of the $\bar{e} W \nu_e$ and $\bar{\mu} W\nu_\mu$ couplings
increases the muon lifetime,
that agrees at about the per-mille level
with the SM prediction obtained from
precision measurements of the electromagnetic coupling and of the $W$ and $Z$ masses.
Assuming that no other new physics beyond the extra operator in eq.~(\ref{eq:nud}) is present,
from a fit of the data in table~\ref{tab:data} we find that
a flavour-universal $\epsilon_{ii}$ is constrained to be $\epsilon_{ii} = (0\pm 0.4)10^{-3}$.
This bound cannot be evaded with flavour non universal
corrections, that are too strongly constrained by
lepton universality tests
in $\tau$ and $\pi$ decays~\cite{Pich}.
In conclusion, $\epsilon_{\mu\mu}$ can possibly generate only a small fraction of the NuTeV anomaly.

In principle, the strong bound from muon decay could be circumvented by
mixing the neutrinos with extra fermions that have
the same $W$ coupling of neutrinos
but a different $Z$ coupling.
In practice, it is not easy to build such  models.

\subsubsection*{Models that only affect the $Z$ couplings}
Only the $Z$ couplings
are modified,  e.g., in models with an
extra U(1) $Z'$ gauge boson that mixes with the $Z$ boson.
The $Z'$ effects can be described by the $ZZ'$ mixing angle, $\theta$,
by the $Z'$ boson mass, $M_{Z'}$, and by the $Z$ gauge current $J_{Z'}$.
At leading order in small $\theta$ and $M_Z/M_{Z'}$,
the tree-level low energy lagrangian gets modified in three ways.
\begin{enumerate}
\item[(1)] the SM prediction for the $Z$ mass gets replaced by
$M_Z^2 =M_Z^{2\rm SM} - \theta^2 M_{Z'}^2$;
\item[(2)] the $Z$ current becomes $J_Z = J_Z^{\rm SM} - \theta J_{Z'}$;
\item[(3)] at low energy, there are the four fermion operators generated by $Z'$ exchange,
beyond the ones generated by the $W_\pm$ and $Z$ bosons:
$$\Lag_{\rm eff}(E \ll M_Z,M_{Z'}) = -\frac{J_{W_+} J_{W_-}}{M_W^2}  -\frac{1}{2} \bigg[ \frac{J_Z^2}{M_Z^2} +
\frac{J_{Z'}^2}{M_{Z'}^2}\bigg] + \cdots.$$
\end{enumerate}
As discussed in section~\ref{oblique},
(1) cannot explain the $\sim 1\%$ NuTeV anomaly.
Here we show that the same happens also for (2):
the $Z$ couplings are constrained by LEP and SLD at the per-mille level,
and less accurately by atomic parity violation data,
as summarized in table~\ref{tab:data}.
However, the less accurate of these data have $\sim 1\%$ errors, and
present  some anomalies.
The $Z\to \nu\bar{\nu}$ rate and
the Forward/Backward asymmetries
of the $b$ and $c$ quarks show a few-$\sigma$ discrepancy with the best-fit SM prediction.
But the $Z\to b\bar{b}$ and $Z\to c\bar{c}$ branching ratios
agree with the SM.
The best SM fit, including also the NuTeV data~\cite{NuTeV},
has $\chi^2 \approx 30$ with 14 d.o.f.
In this situation, it is interesting to study if these anomalies could have a common solution
with $Z$ couplings about $1\%$ different from the SM predictions.
We therefore extract the $Z$ couplings directly from the data,
without imposing the SM predictions for them.
This kind of analysis has a general interest.
Since we are here concerned with the NuTeV anomaly,
we apply our results to compute the range of $(g_L^2, g_R^2)$ consistent with the electroweak data.
We recall that both neutrino and quark couplings enter in determining $g_L^2$ and $g_R^2$.

We assume that the $Z$ couplings are generation universal and $\SU(2)_L$ invariant as in the SM:
we therefore extract from the data the 5 parameters $g_Q$, $g_{U}$, $g_D$, $g_L$ and $g_E$
that describe the $Z$ couplings to the five kinds of
 SM fermions listed in table~\ref{tab:gAi}.
In the context of $Z'$ models, this amounts to assume that the $Z'$ has generation-universal couplings
that respect $\SU(2)_L$.
This  assumption of $\SU(2)_L$ invariance is theoretically well justified,
although one could possibly invent some non minimal model where it does not hold.
On the contrary, the universality assumption only has a pragmatic motivation:
we cannot make a fit with more parameters than data.

We obtain the result shown by the large
red ellipse on the right side of fig.~\ref{fig:plot2}.
This generic class of models gives a best fit value close to the SM prediction.
Although  the error is much larger than in a pure SM fit,
it does not allow to cleanly explain the NuTeV anomaly.
We find that the global $\chi^2$ can be decreased by about 4 with respect to a SM fit:
taking into account that we have five more parameters this is not a statistically significant reduction\footnote{On the contrary,
if we allow for generation universal but non
$\SU(2)_L$-invariant corrections of the $Z$ couplings to $u_L$ and $d_L$ and to $\nu$ and $e_L$,
we get a statistically significant reduction in the global $\chi^2$
($\Delta\chi^2 \approx 21$ with 7 more parameters),
due to the fact that various anomalies, including of course the NuTeV anomaly,
can be explained in this artificial context.
Without including the NuTeV data, the best fit regions in the ($g_L^2, g_R^2$) plane are shifted
towards the NuTeV region.}
in agreement with old similar analyses~\cite{e.g.}.

One could generalize this analysis in several directions.
For example, new physics could shift the on-shell
$Z$ couplings tested at LEP and SLD
differently from the low-energy $Z$ couplings relevant for NuTeV.
Alternatively, there could be flavour dependent
shifts of the $Z$ couplings.
This happens e.g.\ in the model considered in~\cite{Roy},
where it is suggested that the NuTeV anomaly could be reproduced by
a mixing between the $Z$ boson with a $Z'$ boson coupled to the lepton flavour numbers
$L_\mu - L_\tau$.
However, this mixing also shifts the couplings of charged $\tau$ and $\mu$ leptons,
that are too precisely tested by LEP and SLD to allow
for a significant fraction of the NuTeV anomaly.

\section{Loop effects in the MSSM}\label{susy}
It is well known that
supersymmetric contributions to the electroweak precision
observables decouple rapidly. Under the present experimental constraints
it is very difficult to find regions of parameter space where
radiative corrections can exceed a few per-mille. Explaining the NuTeV
anomaly (a 1.2\% discrepancy with the SM prediction for $g_L^2$) with low-energy supersymmetry
looks hopeless from the start. Moreover, the dominant contributions to $\epsilon_1$
in the MSSM are always positive \cite{drees}. It then follows from
eqs.~(\ref{eps}) that, in order to explain at least partially the
measured value of $g_L^2$, the supersymmetric contributions to $\epsilon_3$
should be positive and of ${\cal O}(1\%)$.

\begin{figure}[t]
 \parbox{8cm}{$$
 \includegraphics[width=70mm]{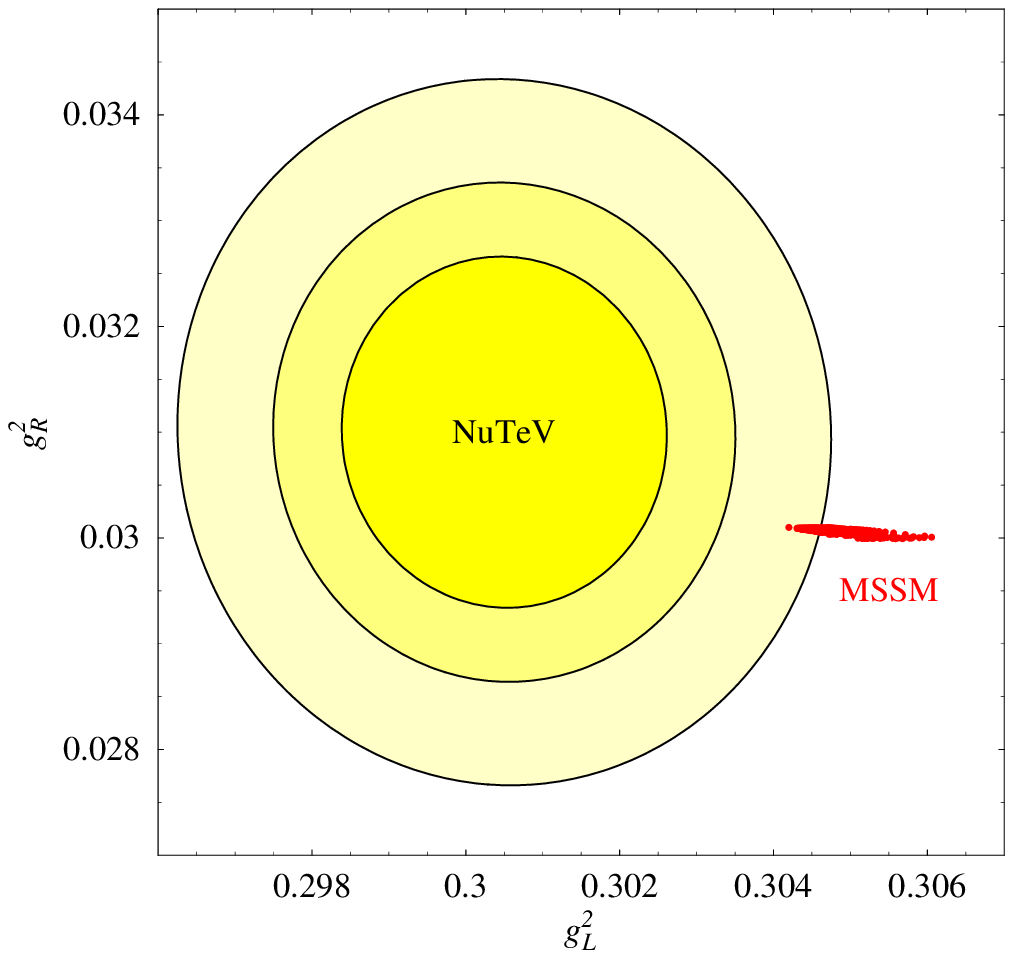} $$
 \caption{\em Shifts in $g_{L,R}^2$ in the supersymmetric scenario
 described in the text (light sleptons and gauginos) for
points in the parameter space
which are not excluded by  experimental constraints.}
 \label{fig:plot4}}\hfill
 \parbox{8cm}{
$$
 \includegraphics[width=70mm]{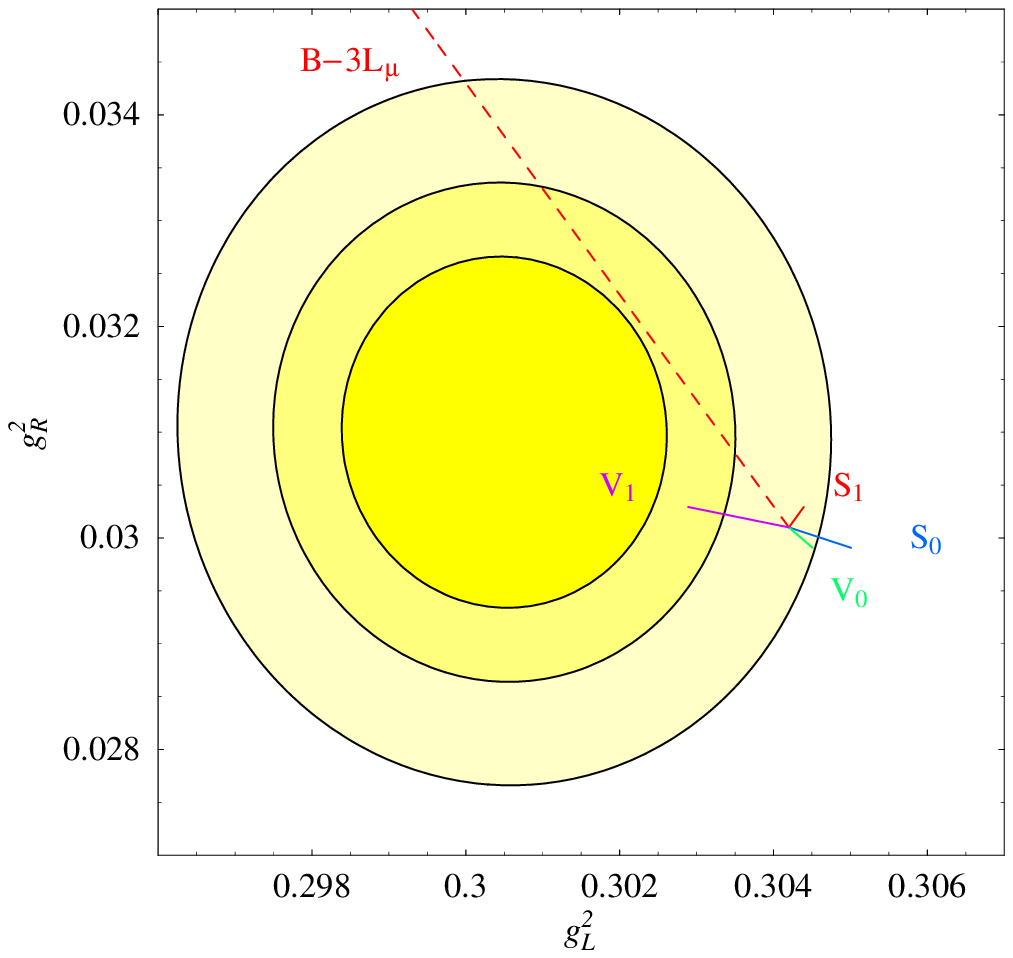} $$
 \caption[]{\em Deviations from the SM prediction that can be induced
by minimal leptoquarks (continuous lines)
and by an extra $B-3L_\mu$ gauge boson (dashed red line)
without conflicting by more than $1\sigma$ with other bounds.
\label{fig:plot3}}}
 \end{figure}

An interesting scenario which can be easily investigated is the one
recently proposed in \cite{altarelli} to improve the global fit
to the electroweak data. As the main contributions of  squark loops
would be a positive shift in $\epsilon_1$, all squarks  can be
assumed heavy, with masses of the order of one TeV.
Relatively large supersymmetric contributions are  then  provided by light
gauginos and sleptons and can be parameterized in terms of only four
supersymmetric parameters ($\tan\beta$, the Higgsino mass $\mu$, the
weak gaugino mass $M_2$, and a supersymmetry-breaking mass of left-handed
sleptons).
The oblique approximation used in section
\ref{oblique} is not appropriate for light superpartners (sneutrinos
can be as light as $50\GeV$). We therefore consider the complete
supersymmetric one-loop corrections in this scenario (see
\cite{altarelli} and refs.\ therein).
Taking into account the various experimental
bounds on the chargino and slepton masses,
we find the potential shifts to
$g_{L,R}^2$ shown in fig.~\ref{fig:plot4}. They
are small and have the wrong
sign. Low-energy supersymmetric loops cannot generate the NuTeV anomaly.

\section{Non renormalizable operators}\label{NRO}
Non renormalizable operators parameterize the
effects of any new physics too heavy to be directly produced.
As discussed in sections~\ref{oblique} and~\ref{couplings}, new physics
that affects the $Z,W_\pm$ propagators or couplings
cannot fit the NuTeV anomaly without some
conflict with other electroweak data.
We now consider
dimension six lepton-lepton-quark-quark operators
that conserve
baryon and lepton number.

We start from a phenomenological perspective, with $\SU(3) \otimes
{\rm U}(1)_{\rm em}$ invariant
four fermion vertices, and determine which vertices
could explain the NuTeV anomaly without conflicting
with other data. Then we next consider which $\SU(2)_L$
invariant operators
generate the desired four-fermion vertices.
In the next sections, we will discuss
new particles whose exchange could
generate these operators.

\medskip

Taking into account Fierz identities,
the most generic Lagrangian that we have to consider can be written as
\begin{eqnarray}\nonumber
\Lag_{\rm eff}  &=& \Lag_{\rm SM}-2 \sqrt{2} G_F\bigg[
\epsilon^{AB}_{ \bar{\ell}_i \ell_j \bar{q}_r q_s}
( \bar{\ell}_i \gamma^{\mu}P_A \ell_j)
(\bar{q}_r \gamma_{\mu}P_B q_s)+\\
&&+
\delta^{AB}_{\bar{\ell}_i \ell_j \bar{q}_r q_s}(\bar\ell_i P_A \ell_j)(\bar q_r P_B q_s)+
t^{AA}_{\bar{\ell}_i \ell_j \bar{q}_r q_s}
(\bar\ell_i \gamma^{\mu\nu} P_A \ell_j)(\bar q_r  \gamma_{\mu\nu}  P_A q_s)
\bigg].
\label{epsilon}
\end{eqnarray}
where $\gamma_{\mu\nu}=\frac{i}{2}[\gamma_\mu , \gamma_\nu]$,
$P_{R,L}\equiv (1\pm \gamma_5)/2$ are the
right- and left-handed projectors,
$q$ and $\ell$ are
any quark or lepton, $A,B=\{L,R\}$
and $\epsilon$, $\delta$ and $t$ are dimensionless coefficients.
In order to explain the NuTeV anomaly, new physics
should give a negative
contribution to $g_L^2$.
This can be accomplished by
\begin{enumerate}
\item reducing the NC $\nu_\mu$-nucleon cross section.
The operators given by new physics must
contain left-handed first generation quarks.

\item increasing the CC $\nu_\mu$-nucleon cross section.
The quarks do not need to be left-handed, and the quark  in the
final state  needs not to be of the  first generation.
\end{enumerate}

We now show that `scalar' operators (the ones with coefficients $\delta$)
cannot explain NuTeV,
left-handed `vector' operators (with coefficient $\epsilon^{LL}$)
can realize the first possibility and
`tensor' operators (with coefficient $t$) perhaps the second one.

The {\bf scalar operators} with coefficient $\delta$ contribute
to the charged current.
In order to accommodate the NuTeV anomaly, these operators
should appear with a relatively large coefficient  $\delta \circa{>} 0.1$,
since their contribution  to CC scattering has only a negligible interference with the dominant SM amplitude.
The interference is suppressed by fermion masses (for first generation
quarks) or by CKM mixings (if a quark of higher generation is involved).
For first generation quarks, this value of $\delta_{ \bar{\mu} \nu_\mu \bar{u} d }$ is inconsistent with $R_\pi$. When new physics-SM interference
is included in this ratio, it becomes~\cite{Shanker:1982nd}
\begin{equation}
R_\pi\equiv  \frac{\hbox{BR} (\pi\to e \bar{\nu}_e)}{\hbox{BR}(\pi\to \mu \bar{\nu}_\mu)}  = R_\pi^{\rm SM}
  \left[ 1 - 2 (\epsilon^{LL}_{ \bar{\mu} \nu_\mu \bar{u} d }- \epsilon^{LL}_{ \bar{e} \nu_e \bar{u} d }) -
\frac{2m_\pi^2}{ m_\mu(m_u +m_d)}
\delta^{LP}_{ \bar{\mu} \nu_\mu \bar{u} d } \right].
\label{pi}
\end{equation}
The measured value,
$R_\pi = (1.230
\pm 0.004)
\times 10^{-4}$ \cite{PDG}, agrees with the SM
prediction~\cite{Marciano:1993sh,Finkemeier:1996gi}
$$  R_\pi^{\rm SM}
=  \frac{m_e^2 ( m_\pi^2 - m_e^2)^2}{m_\mu^2 ( m_\pi^2
- m_\mu^2)^2}(1-16.2 \frac{\alpha_{\rm em}}{\pi})$$
which implies $\delta_{ \bar{\mu} \nu_\mu \bar{u} d }\lsim 10^{-4}$.
Furthermore, scalar operators
which produce a  $s,c$ or $b$ quark in the final state
also cannot explain the NuTeV anomaly. The
values of $\delta$ required would be in conflict
with upper bounds on FCNC meson decays such as
$K^+ \rightarrow \pi^+ \mu \bar{\mu}$,
$D^+ \rightarrow \pi^+ \mu \bar{\mu}$, and
$B^0 \rightarrow  \mu \bar{\mu}$.

\bigskip

{\bf Vector operators} can possibly generate the NuTeV
anomaly if they are of $LL$ type.
Assuming first generation quarks,
the operators in eq.~(\ref{epsilon}) shift
$g_L^2$ as
 \bea
\delta g_L^2 &=&
   2 \big[    g_{Lu}~
\epsilon^{LL}_{ \bar{\nu}_\mu \nu_\mu \bar{u} u} +
   g_{Ld}~
\epsilon^{LL}_{ \bar{\nu}_\mu \nu_\mu \bar{d} d}
 - g_L^2 ~\epsilon^{LL}_{ \bar{\nu}_\mu \mu \bar{d} u}  \big].
 \label{DgL2}
\eea
The CC term, $\epsilon^{LL}_{ \bar{\nu}_\mu \mu \bar{d} u}$, cannot
alone fit the
NuTeV anomaly without overcontributing to the $\pi\to \mu\bar{\nu}_\mu$ decay.
In principle, one could allow for cancellations between different
contributions to $R_\pi$ in eq.~(\ref{pi}).
However LEP~\cite{LEP2combined} (and bounds from atomic parity violation~\cite{LEPEWWG})
exclude the simplest possibility,
$\epsilon^{LL}_{ \bar{\mu} \nu_\mu \bar{u} d }=\epsilon^{LL}_{ \bar{e} \nu_e \bar{u} d}$.

We now assume that these vector operators are generated by new physics heavier than
the maximal energy of present colliders (about few hundred GeV),
and study the bounds from collider data.
Operators  involving second generation
leptons are
constrained by the Tevatron; LEP and HERA are not
sensitive to them\footnote{Both at NuTeV and at CCFR these operators
may induce deviations from the SM values of $g_L,g_R$.
Within their errors, the CCFR data
are consistent with both the SM and with NuTeV,
as clearly shown in fig.~3 of~\cite{CCFR}.
Therefore CCFR bounds on non renormalizable
operators, reported by the PDG~\cite{PDG}, cannot conflict with NuTeV.}.
In the case of vector operators,
the Tevatron sets a limit~\cite{TevatronContact}
$|\epsilon^{LL}_{\bar{\mu} \mu \bar{q} q}|\circa{<}0.03$
($q=u$ is slightly more constrained than $q=d$,
because protons contain more $u$ than $d$).
Assuming SU(2)$_L$-invariance (that relates the
$\epsilon^{LL}_{\bar{\ell}\ell \bar{q} q}$
with $\ell = \{\mu,\nu_\mu\}$ and $q=\{u,d\}$, see below)
the Tevatron bound is close but consistent with the value suggested by NuTeV,
$| \epsilon^{LL}_{\bar{\nu}_\mu\nu_\mu \bar{q} q} | \sim 0.01$.



\bigskip

{\bf Tensor operators} could possibly produce
the NuTeV anomaly via mechanism 2,
because $\pi$-decays give no bound on
$t_{ \bar{\mu} \nu_\mu \bar{u} d }$
(using only the $\pi$ momentum it is not
possible to write any antisymmetric tensor).
Tensor operators have not
been studied in~\cite{TevatronContact},
but if they are generated by physics at a scale $\gg M_Z$,
the value of $t_{ \bar{\mu} \nu_\mu \bar{u} d }\sim 0.1$
necessary to fit the NuTeV anomaly is within (and probably above)
the sensitivity of present Tevatron data.
Furthermore we do not know how
new physics (e.g.\ exchange of new scalar or vector particles~\cite{Pich:1995vj})
could generate only tensor operators,
without also generating the scalar operators
that overcontribute to $R_\pi$.
We will therefore focus on vector operators.


\bigskip

We now consider $\SU(2)_L$-invariant operators.
We have shown  that the
NuTeV anomaly could be explained by the four-fermion
vertex $ ( \bar{\nu}_\mu \gamma^{\mu}P_L \nu_\mu)
(\bar{q}_1 \gamma_{\mu}P_L q_1)$.
Only two $\SU(2)_L$ invariant operators operators can generate this vertex
$$
{\cal O}_{LQ} =[\bar{L} \gamma_\mu L][\bar{Q} \gamma_\mu Q], \qquad
{\cal O}_{LQ}' = [\bar{L} \gamma_\mu  \tau^a L][\bar{Q} \gamma_\mu \tau^a Q].$$
We left implicit  the $\SU(2)_L$ indices, on which the Pauli matrices $\tau^a$ act.
Other possible 4 fermion operators, with different
contractions of the $\SU(2)_L$ indices,
can be rewritten as linear combinations of these two operators.

The  NuTeV anomaly can be fit by
${\cal O}_{LQ}$
if it is present in $\Lag_{\rm eff}$ as
$ (-0.024\pm 0.009)\, 2\sqrt{2} G_{\rm F}{\cal O}_{LQ}$,
as discussed above.
The operator
$${\cal O}'_{LQ}= [\bar{\nu}_\mu \gamma_\mu \nu_\mu
-\bar{\mu}_L\gamma_\mu \mu_L][\bar{u}_L \gamma_\mu u_L -
\bar{d}_L
\gamma_\mu d_L] + 2[\bar{\mu}_L \gamma_\mu\nu_\mu][\bar{u}_L \gamma_\mu
d_L]+2[\bar{\nu}_\mu\gamma_\mu \mu_L][\bar{d}_L\gamma_\mu u_L]$$
also can fit the NuTeV anomaly.
However its CC part overcontributes to $\pi\to \mu\bar{\nu}_\mu$,
giving a contribution to $\epsilon^{LL}_{ \bar{\mu} \nu_\mu \bar{u} d } $
about 10 times larger than what allowed by $R_\pi$, see eq.~(\ref{pi}).

These  operators could be induced e.g.\ by leptoquark or
$Z'$ boson exchange, which we study in the
following two sections.  A critical difference between these
possibilities
is that leptoquarks must
be heavier than  about $200\GeV$~\cite{Abbott:2000ka,Abe:1998it,oggi},
whereas a neutral $Z'$ boson could also
be lighter than about $10 \GeV$ (see section~\ref{Z'}).
Leptoquarks are charged
and coloured particles that would be pair-produced
at colliders, if kinematically possible.
If the NuTeV anomaly is due to leptoquarks,
their effects should be seen at run II of the Tevatron or at the LHC.
If instead the NuTeV anomaly were due to a weakly coupled light $Z'$,
it will not show up at Tevatron or LHC.

\renewcommand{\arraystretch}{1.5}
\begin{table}
$$\begin{array}{clcccccc}
\hbox{LQ} &
\multicolumn{1}{c}{\Lag_{\rm eff}}& \delta g_L^2
& \epsilon_{\bar{\nu}_\mu\nu_\mu\bar{d}{d}}^{LL}
&\epsilon^{LL}_{\bar{\mu}\mu\bar{u}{u}}
& \epsilon_{\bar{\nu}_\mu\nu_\mu\bar{u}{u}}^{LL}
&\epsilon^{LL}_{\bar{\mu}\mu\bar{d}{d}}
& \epsilon_{\bar{\nu}_\mu \mu\bar{d}{u}}^{LL},\epsilon^{LL}_{\bar{\mu}\nu_\mu\bar{u}{d}}
\cr
\hline\hline S_0  &
\phantom{+}\frac{|\lambda|^2}{4m^2}({\cal O}_{LQ} - {\cal O}'_{LQ}) &
 0.12 \alpha&
- \alpha/2 &- \alpha/2 & 0 &  0 &  \alpha/2 \\ \hline
& & & 0 & 0 & - \alpha & 0 & 0 \\
S_1  &
\phantom{+}\frac{|\lambda|^2}{4m^2}({\cal O}'_{LQ} +3 {\cal
O}_{LQ}) &
0.03 \alpha& - \alpha/2 & - \alpha/2 & 0 & 0  & - \alpha/2\\
& & & 0 & 0 & 0 & - \alpha & 0 \\ \hline
V_0 &
-\frac{|\lambda|^2}{2m^2}({\cal O}'_{LQ} + {\cal O}_{LQ}) &
0.09 \alpha &
0 & 0 &  \alpha &  \alpha &  \alpha \\ \hline
& & & 0 &  2 \alpha & 0 & 0 & 0 \\
V_1  &
\phantom{+} \frac{|\lambda|^2}{2m^2}({\cal O}'_{LQ} - 3{\cal O}_{LQ} )
 &-0.40\alpha
 & 0 & 0  &  \alpha &  \alpha & -  \alpha \\
& & &  2 \alpha & 0 & 0 & 0 & 0
\end{array}$$
\caption{\em The four leptoquarks that generate the operators suggested
by the NuTeV anomaly. Columns two and three
are  the effective Lagrangian and the
contribution to $g_L^2$
(in terms of $\alpha
\equiv  |\lambda^2|/2\sqrt{2} G_{\rm F}m^2$)
for leptoquarks with $\SU(2)_L$ degenerate masses.
In the last columns, we give the coefficients of
the individual four fermion operators,
separating the  triplet members.
\label{tab:LQ}}
\end{table}

\section{Leptoquarks}\label{LQ}
Leptoquarks are scalar or vector bosons with
a coupling to leptons and quarks. In this section, we
consider leptoquarks which induce baryon and lepton number
conserving four-fermion vertices.

The symmetries of the SM allow  different types of leptoquarks,
which are listed in~\cite{LQ}. There are four
leptoquarks that couple to $Q L$, so these are candidates
to   explain the NuTeV anomaly.  They are
the scalar $\SU(2)_L$ singlet ($S_0$) and triplet ($S_1^a$), and
the vector $\SU(2)_L$ singlet ($V_{0\mu}$)  and triplet
($ V_{1\mu}^a$), with interaction Lagrangian
\beq
 \lambda_{S_0} [QL] S_0 +
 \lambda_{S_1} [Q\tau^a L] S_1^a +
 \lambda_{V_{0}} [\bar{Q} \gamma_\mu L] V_{0\mu} +
 \lambda_{V_{1}} [\bar{Q} \gamma_\mu \tau^a L] V_{1\mu}^a + \hbox{h.c.}
\eeq
We do not speculate on how the above leptoquarks
could arise in specific models.

\medskip

Consider first the scalar $S_0$.
The lower bound on leptoquark masses from
the Tevatron  is
$200\GeV$~\cite{Abbott:2000ka,Abe:1998it},
therefore at NuTeV  leptoquarks are
equivalent to effective operators.
Tree level exchange of $S_0$,
with mass $m$ and coupling  $\lambda  [Q_1 L_2]S_0$ (1 and 2 are
generation indices),
induces the  four-fermion operator
$$\Lag_{\rm eff} =\frac{|\lambda|^2}
{m^2} (Q_1L_2)(Q_1{L}_2)^{\dagger}  =  \frac{|\lambda|^2}
{4m^2}({\cal O}_{LQ} - {\cal
O}'_{LQ}),
$$
The sign of the
operator is fixed, because the coupling
constant is squared.
We see that $S_0$ cannot explain the NuTeV anomaly, because it
generates ${\cal O}_{LQ}$ with the wrong sign (it gives a
positive contribution to $g_L^2$),
and because it also generates the
unwanted operator ${\cal O}'_{LQ}$.

In the context of supersymmetric models without $R$-parity
$S_0$ can be identified with a
$\tilde{D}^c_g$ squark of generation $g$ and
superpotential interaction $\lambda'_{2g1} L_2 D^c_g Q_1$.
It is interesting to explore further the possible contributions
of $R$-parity violating squarks at NuTeV.
In supersymmetric models, $\tilde{D}^c$ is accompanied by a scalar
$\SU(2)_L$ doublet squark (leptoquark), $\tilde{Q}$.
The exchange of $\tilde{Q}$
only modifies the right-handed coupling $g_R$, so that it cannot
explain NuTeV by itself.
Mixing of right- and left-handed squarks
generates dimension seven
operators.
This mixing is usually, but not always, negligibly small
(e.g.\ one can consider large $\tan\beta$, or non minimal models).
The relevant $\Delta L=2$ four-fermion operators are
\beq
\lambda'_{ijk} \lambda'_{mnj}
[\bar{d}_k P_L \nu_i][ \bar{Q^c_n} P_L L_m]=
\lambda'_{ijk} \lambda'_{mnj}
[\bar{d}_k P_L \nu_i] [\bar{u}^c_n P_L e_m - \bar{d}^c_n P_L \nu_m]
\eeq
These operators cannot account for the NuTeV anomaly:
they do not interfere with the SM amplitude and contribute to both NC
and CC, leading to a positive correction to $g_L$.

\medskip

In table~\ref{tab:LQ}, we list
the effective four-fermion operators, and the contribution
to $g_L^2$, of $S_0$,  $S_1$, $V_0$
and $V_1$. In the $\Lag_{\rm eff}$ column
we have assumed that the members of
triplet leptoquarks are degenerate.
Only the vector $\SU(2)_L$
 triplet leptoquark
 gives a negative contribution to  $g_L^2$.
In all cases ${\cal O}_{LQ}$ is generated together with the unwanted
${\cal O}'_{LQ}$ operator, that
overcontributes to the $\pi\to \mu\bar{\nu}_\mu$ decay,
as discussed in the previous section.
These features are also shown in fig.~\ref{fig:plot3}, where we plot
the deviations from the SM prediction induced by the $S_0$, $S_1$, $V_0$, $V_1$
leptoquarks imposing that they should not overcontribute
to $R_\pi$ by more than $1\sigma$.

In the subsequent columns of table~\ref{tab:LQ} we generalize the effective Lagrangian
assuming that $\SU(2)_L$ breaking effects
split the triplets in a general way.
In this situation, the scalar and vector
$\SU(2)_L$ triplet leptoquarks can explain the NuTeV anomaly.
In the scalar (vector) case, NuTeV can be fit by
reducing  the mass of  the triplet member that induces
$\epsilon^{LL}_{ \bar{\nu}_2  \nu_2 \bar{u} u}$
($\epsilon^{LL}_{ \bar{\nu}_2  \nu_2 \bar{d} d}$),
by a factor of $\sqrt{2 } $.
From~\cite{rho}, we expect that such split multiplets
are consistent with precision electroweak measurements.

We conclude that the NuTeV anomaly
cannot be generated by $\SU(2)_L$
 singlet or doublet leptoquarks,
or by  triplet leptoquarks with degenerate masses.
However, triplet leptoquarks with carefully chosen
mass splittings between the triplet members can
fit the NuTeV data --- and this explanation should be tested
at Run II of the Tevatron or at LHC.

\section{Unmixed extra U(1) $Z'$ boson}\label{Z'}
The sign of the dimension 6 lepton/quark operators generated by an extra $Z'$ vector boson
depends on the lepton and quark
charges under the ${\rm U}(1)'$ gauge symmetry.
Therefore, with generic charges, it is possible to generate a correction to neutrino/nucleon scattering
with the sign suggested by the NuTeV anomaly.
In order to focus on theoretically appealing $Z'$ bosons,
we require that
\begin{itemize}
\item Quark and lepton mass terms are neutral under the extra ${\rm U}(1)'$.
We make this assumption because
experimental bounds on flavour and CP-violating processes
 suggest that we do not have a flavour symmetry at the electroweak scale.

\item The $Z'$ couples to leptons of only second generation.
Bounds from (mainly) LEP2~\cite{LEP2combined} and older $e\bar{e}$ colliders
would prevent to
explain the NuTeV anomaly in presence of couplings to first generation leptons.
We avoid couplings to third generation
leptons just for simplicity.

\item The extra ${\rm U}(1)'$ does not have anomalies that require extra light fermions
charged under the SM gauge group.

\end{itemize}
The only gauge symmetry that satisfies these conditions is $B-3L_\mu$ (for related work see~\cite{chang}),
where
$B$ is the baryon number and
$L_\mu$ is the muon number.\footnote{The neutrino masses suggested by
oscillation data~\cite{SKatm} do not respect $L_\mu$. If we allow third generation leptons to be
charged under the extra ${\rm U}(1)'$ symmetry, we could have a $B- c L_\mu - (3-c)L_\tau$
gauge group ($c$ is a constant). However even a $L_\mu \pm
L_\tau$ symmetry would not force a successful pattern of
neutrino masses and mixings,
that rather suggest a flavour symmetry containing $L_e - L_\mu - L_\tau$~\cite{numasses}.}
Under these restrictions, the sign of the $Z'$ correction to neutrino/nucleon scattering is fixed,
and this $Z'$ allows to fit the NuTeV anomaly.
In fact, the four-fermions operators generated by $Z'$ exchange are
\begin{eqnarray*}
\Lag_{Z'} &=& -\frac{g_{Z'}^2}{2(M_{Z'}^2-t)}\bigg[\bar{Q}\gamma_\mu Q -\bar{U}^c \gamma_\mu U^c
-\bar{D}^c\gamma_\mu D^c -9
 \bar{L}_2 \gamma_\mu L_2+9\bar{E}_\mu^c \gamma_\mu E_\mu^c\bigg]^2
\end{eqnarray*}
where $t$ is momentum transferred: $t \sim - 20\GeV^2 $ at NuTeV.
The best fit of the NuTeV anomaly is obtained for
(see fig.~\ref{fig:plot3})
\begin{equation}\label{eq:Z'x}
\sqrt{M_{Z'}^2-t}\approx  g_{Z'}\,3\TeV.
\end{equation}
We now discuss the experimental bounds on such a $Z'$.

\subsubsection*{Collider bounds}
The bounds from Tevatron~\cite{TevatronZ'}
$$\sigma(p\bar{p}\to Z'X\hbox{ at $\sqrt{s} = 1.8\TeV$})\hbox{BR}(Z'\to
\mu\bar{\mu}) < 40\,\hbox{fb}\quad(95\%\hbox{CL})$$
and LEP~\cite{LEP1}\footnote{The total number of measured
$Z\to\mu\bar{\mu}\mu\bar{\mu}$ agrees
with the SM prediction and there is no peak in the $\mu\bar{\mu}$
invariant mass.
In order to extract a precise bound on the $Z'$ mass from these data
one should take into account the experimental cuts and resolution.
}
$$\hbox{BR}(Z\to \mu\bar{\mu}Z')\hbox{BR}(Z'\to \mu\bar{\mu})\circa{<}
\hbox{few}\times 10^{-6}$$
imply  that $M_{Z'}$ cannot be comparable to $M_Z$.
One needs either a light $Z'$,
$M_{Z'}\circa{<} 10 \GeV$, or a heavy $Z'$, $M_{Z'}\circa{>}
600\GeV$~\cite{TevatronZ'}.
Perturbativity implies $M_{Z'}\circa{<} 5\TeV$.

\subsubsection*{The anomalous magnetic moment of the muon}
The $Z'$ gives a correction to the anomalous magnetic moment of the muon.
Assuming $M_{Z'}\gg m_\mu$, we get
$$  a_\mu = a_\mu^{\rm SM} +\frac{27g_{Z'}^2}{4\pi^2}\frac{m_\mu^2}{M_{Z'}^2} = a_\mu^{\rm SM} +8.4~10^{-10} \bigg(\frac{3\TeV}{M_{Z'}/g_{Z'}}\bigg)^2.$$
At the moment, the $a_\mu$ measured by the $g-2$ collaboration~\cite{g-2}
is slightly higher than the SM prediction,
$a_\mu^{\rm exp}- a_\mu^{\rm SM} = (20\pm 18)10^{-10}$,
if one employs $a_\mu^{\rm had} = (697\pm 10)10^{-10}$
for the hadronic polarization contribution~\cite{had} and
$a_\mu^{\rm lbl} = (9\pm 2)\,10^{-10}$~for the
light-by-light contribution~\cite{lbl}.
One could still prefer to estimate it as $a_\mu^{\rm lbl} = (0\pm 10)\,10^{-10}$,
obtaining a larger discrepancy and error.

Therefore, if $M_{Z'}/g_{Z'}\approx 3\TeV$, as suggested by the NuTeV anomaly
in the heavy $Z'$ case,
the $Z'$ correction to the $g-2$ is comparable to the sensitivity of present experiments.
If instead $M_{Z'} \circa{<} 5\GeV$, the $Z'$
that fits the NuTeV anomaly gives a larger correction to
the muon $g-2$, see eq.~(\ref{eq:Z'x}).
For example, for $M_{Z'}\approx 3\GeV$ one can fit the central value of
 $a_\mu^{\rm exp} - a_\mu^{\rm SM}$.
On the other hand, a
 $Z'$ lighter than $(1\div2)\GeV$ cannot explain the NuTeV anomaly
without overcontributing to $a_\mu$.
Similar light $Z'$ models were proposed~\cite{russi}
as a possible source of the discrepancy
between $a_\mu^{\rm exp}$ and previous  SM computations~\cite{g-2}.

\subsubsection*{Other bounds}
Quantum corrections generate an unwanted kinetic mixing between
the $Z'$ and the SM hypercharge boson~\cite{Holdom}.
A light $Z'$ needs a small gauge coupling $g_{Z'}$, making these quantum effects negligible.

The $Z'$ could contribute to the decay
of $q \bar{q}$ mesons into $\bar{\mu} \mu$. This
is negligibly small for $g_{Z'} \sim  M_{Z'}/ 3\TeV$,
unless $m_{Z'}$ is very close to the meson mass $m_{q \bar{q}}$. There
are various $c \bar{c}$ mesons in the few GeV mass range,
but $\Gamma_{Z'}$ is very narrow, so the $Z'$ is only
ruled out in narrow windows $m_{Z'} \approx  m_{q \bar{q}}
\pm  \Gamma_{{q \bar{q}}}$~\cite{Bailey:1995qv}.

The $Z'$ that can fit the NuTeV anomaly does not give
significant corrections to rare $K$, $D$ and $B$ decays.
Let us consider for example the $K^+\to \pi^+ \nu\bar{\nu}$ decay.
This is a sensitive probe because
the dominant SM $Z$ penguins that know that $\SU(2)_L$ is broken (and can therefore
generate the FCNC vertex $m_t^2 [\bar{s}_L \gamma_\mu Z_\mu d_L]$,
with GIM cancellations spoiled by the large top mass) are suppressed by
the small mixing $V_{ts}^*V_{td}$.
On the contrary, penguin loops of quarks that do not know that ${\rm U}(1)'$ is broken
(that therefore only generate the
$q^2[\bar{s}_L \gamma_\mu Z'_\mu d_L]$ operator,
where $q$ is the $Z'$ momentum and the GIM cancellations
is only spoiled by logarithms of quark masses)
are suppressed only by the larger Cabibbo mixing $V_{cd}$.
The $Z'$ suggested by NuTeV gives a negligible correction even to
this favorable $K^+\to \pi^+\nu\bar{\nu}$ decay.\footnote{A $Z'$ boson
with mass $M_{Z'}\sim
100\MeV$ (this case is not motivated by NuTeV)
 would mediate the resonant decay $K^+\to \pi^+ Z'\to \pi^+ \nu \bar{\nu}$, producing an
excess of monoenergetic $\pi^+$ in the $K^+$ rest frame,
compatibly with the first experimental data~\cite{Kpinuexp}.}


\medskip

Ref.~\cite{dimuon} claims a statistically significant hint of dimuon events
(three events, versus an estimated background of $0.07\pm0.01$ events),
possibly generated by some neutral long lived
particle with production cross section $\sim 10^{-10}/\GeV^2$
of few GeV mass (see also \cite{dimuonint}).
The $Z'$ suggested by NuTeV  can have the right mass and cross section, but
is not sufficiently long lived. However,
it could  partially decay in sufficiently long lived
neutral fermions. Extra light neutral fermions are required to cancel
gravitational anomalies and ${\rm U}(1)^{\prime 3}$ anomalies.

\subsubsection*{$Z'$ burst and the GZK cutoff}
The $Z'$ gives a narrow resonant contribution to $\nu\bar{\nu}$
scattering,
which could perhaps generate ultrahigh energy cosmic ray
events with $E \sim 10^{20}$ GeV.
The analogous $Z$ resonance~\cite{ringwald}
has been considered as a possible source of
the observed events above the
Greisen-Zatsepin-Kuzmin (GZK) cutoff~\cite{GZK}.
A cosmic ray neutrino that scatters with
a nonrelativistic cosmic microwave background neutrino
would encounter the $Z'$ resonance at the energy
\beq
E_\nu^{Z'\,\rm res} = \frac{M_{Z'}^2}{2 m_\nu} = 10^{20}\eV
\bigg(\frac{M_{Z'}}{3\GeV}\bigg)^2\frac{0.05\eV}{m_\nu}
\label{CR}
\eeq
where we have used a neutrino mass
suggested by atmospheric oscillation data~\cite{SKatm}.
A resonance at a $Z'$ mass
of few GeV is more suitable than a resonance on
the $Z$, where a larger incident neutrino energy
$E_\nu^{Z\,\rm res} =M_Z^2/2m_\nu \circa{>} 4 \times 10^{21}$ eV would be
required, even if
neutrinos are as heavy as possible, $m_\nu\circa{<} 1\eV$.
The $Z$ burst scenario is problematic because
it seems difficult to imagine
a cosmological source that produces enough very energetic neutrinos without
producing, at the same time, too many photons.

Although the $Z'$ requires less energetic cosmic neutrinos than the $Z$,
roughly the same total flux~\cite{ringwald} is required in the two cases,
because the $Z$ and the $Z'$ have
comparable energy-averaged cross sections. In fact
the NuTeV and $g-2$ data suggest
$\Gamma_{Z'}/M_{Z'}\sim \Gamma_Z/M_Z$ and
$$\sigma(\nu\bar{\nu}\to Z'\to f)
\approx  \frac{12\pi^2\Gamma_{Z'}}{M_{Z'}}
\hbox{BR}(Z'\to f) \hbox{BR}(Z'\to \nu\bar{\nu}) \delta(s-M_{Z'}^2)$$
where $f$ is any final state.
In conclusion, a $Z'$ burst could generate the
observed cosmic rays above the GZK cutoff more easily than the $Z$ burst.

\section{Summary}
We have studied the possible origin of the NuTeV anomaly.
Our main results are:
\begin{itemize}

\item {\bf QCD effects}.
Because of  the approximate use of the Paschos-Wolfenstein relation,
the discrepancy between the NuTeV result and the SM prediction is fairly
independent of which set of standard parton distribution functions is
employed. This is no longer true if one drops some of the
simplifying assumptions which are usually made  in the PDF fits.
A small asymmetry between the momentum carried by
strange and antistrange quarks in the nucleon,
suggested by a recent analysis of neutrino data~\cite{BPZ},
could explain  half of the discrepancy between NuTeV and the SM.
Such an asymmetry has been set to zero in more recent global parton sets,
but the value suggested in~\cite{BPZ} seems  compatible with the data used in these
fits.
It would be desirable to update the analysis of~\cite{BPZ}
including  more recent di-muon  data from NuTeV \cite{Goncharov},
though settling the issue is likely to require more
detailed information, such as it could be achieved at a neutrino
factory~\cite{nufact}. A tiny violation of isospin symmetry of parton
distributions, largely compatible with current data, would have a similar
impact. Both these effects may have to be taken into account in the
evaluation of the systematic error.

\item {\bf
Generic corrections to the propagators or couplings of the SM gauge bosons}
can only produce
a small fraction of the NuTeV anomaly, as shown in fig.~\ref{fig:plot2}.
In order to perform such a general analysis we have
extracted the `oblique' parameters and the SM gauge couplings directly from
a fit of precision data, without imposing the SM predictions.
We have assumed that the $Z$ couplings are generation universal and respect $\SU(2)_L$, as in the SM.
In principle, the NuTeV anomaly could be explained by new physics that only shifts the $\bar{\nu} Z \nu$
couplings. However this situation is not realized by mixing the $Z$ with extra vector bosons,
nor by mixing the neutrino with extra fermions.

\item {\bf MSSM.}
Loop corrections in the MSSM have generally the wrong sign and are far
too small to contribute significantly to the NuTeV observables.

\item {\bf Contact operators}.
Dimension six quark-quark-lepton-lepton operators can fit
  the NuTeV anomaly consistently with other data.
  The desired operators are neutral current, left-handed
  four fermion vertices of the form
  $ \eta (\bar{\nu}_{\mu} \gamma^{\alpha} {\nu}_{\mu})
  ( \bar{q} \gamma_{\alpha} P_L q)$, where $q = \{u,d\}$.
  The coefficient $\eta$ must be of order $0.01 \times 2 \sqrt{2}
  G_F$, and the sign is fixed by requiring
a negative interference with the SM.
Effects of these operators should be seen at run II of Tevatron,
unless they are generated by very  weakly coupled light particles.
If one restricts the analysis to $\SU(2)_L$-invariant operators,
only the operator $[\bar{Q}_1 \gamma^\alpha Q_1][\bar{L}_2 \gamma_\alpha L_2]$
can fit NuTeV.

\item  {\bf Leptoquarks}.
$\SU(2)_L$ singlet and triplet leptoquarks can induce
  these operators --- but if the leptoquark masses
  are $\SU(2)_L$ degenerate, either the sign is wrong
  or other unacceptable operators are also generated.
A $\SU(2)_L$ triplet leptoquark of spin one
 is a partial exception (see fig.~\ref{fig:plot3})
at least from a purely phenomenological perspective.
  Non degenerate triplet leptoquarks could fit the  NuTeV results, but
  squarks in $R$-parity violating supersymmetry cannot.

\item {\bf Extra U(1) vector bosons}.
A $Z'$ boson that does not mix with the $Z$ boson
can generate the
NuTeV anomaly
(see fig.~\ref{fig:plot3}),
if its gauge group is $B-3L_\mu$, the minimal choice suggested by
theoretical and experimental inputs.
The $Z'$ can be either heavy, $600\GeV \circa{<}M_{Z'}\circa{<} 5\TeV$, or light,
$1\GeV\circa{<}M_{Z'}\circa{<}10\GeV$. The  $Z'$ that fits the NuTeV anomaly also increases the
muon
$g-2$ by $\sim 10^{-9}$
and (if light) gives  a  $Z'$ burst to
cosmic rays just above the GZK cutoff without requiring neutrino masses heavier
that what suggested by oscillation data.

\end{itemize}

\paragraph{Acknowledgments}
We thank  E.~Barone, G.~Giudice, G.~Isidori, M. Mangano,
K.~McFarland,
R.G.~Roberts, A.~Rossi and I.~Tkachev for useful
discussions. A.S. thanks R.~Rattazzi for his insight into
ionization corrections to the $\mu^+$ range in iron.
This work is partially supported by Spanish MCyT grants
PB98-0693 and FPA2001-3031, by the Generalitat Valenciana
under grant GV99-3-1-01, by the TMR network
contract HPRN-CT-2000-00148 of the European Union,
and by EU TMR  contract FMRX-CT98-0194 (DG 12-MIHT).
S.D.\ is supported by the Spanish Ministry of Education in the program
``Estancias de Doctores y Tecn\'ologos Extranjeros en Espa\~na''
and P.G.\ by a EU Marie Curie Fellowship.

 \paragraph{Note added}
 The $\nu_\mu$ ($\bar{\nu}_\mu$)
 NuTeV beam contains a $1.7\%$ contamination of $\nu_e$ ($\bar{\nu}_e$).
 A recent paper~\cite{Giunti} suggests an explanation of the NuTeV
 anomaly
 assuming that $20\%$ of $\nu_e,\bar{\nu}_e$ oscillate into sterile
 neutrinos.
 The suggested oscillation parameters
 are not consistent with disappearance experiments (unless {\sc Bugey}
 and {\sc  Chooz} underestimated the theoretical error on their reactor
 fluxes).
 Furthermore, NuTeV not only predicts the $\nu_e$ and $\bar{\nu}_e$
 fluxes
 through a MonteCarlo simulation, but also  measures them
 directly (see pag.\ 26 of the transparencies in~\cite{NuTeV}).
 The agreement between the two determinations, at the few $\%$ level,
 contradicts the oscillation interpretation.
 
 \paragraph{Note added after publication (June 2002)}
 
 In a recent publication \cite{new},
 the NuTeV collaboration investigated the effect of a strange quark
 asymmetry and of isospin violation on their electroweak result.
 Specifically, they claim that the strange quark asymmetry is severely
 constrained by the dimuon data of Ref.~\cite{Goncharov}, and that
 the effect of strange quark asymmetry and isospin violation might be
 considerably diluted due to the fact that NuTeV does not measure
 directly the Paschos-Wolfenstein ratio $R_{\rm PW}$, Eq.~(\ref{eq:PW}).
 
 The claim that the dimuon (i.e.\ $\nu s\to \mu c$ scattering)
 data~\cite{Goncharov,new} provides evidence against the
 strange asymmetry $s^-\approx +2~10^{-3}$ of the BPZ global
fit~\cite{BPZ}
 (and in fact suggest a {\em negative} asymmetry
 $s^{-} = -(2.7\pm 1.3)10^{-3}$~\cite{Goncharov,new}) appears dubious, for the following
 reasons:
 
 \begin{enumerate}

 \item The parametrization of
 the strange and antistrange distributions assumed by NuTeV is
 unphysical, in that it violates the constraint that the proton and
 neutron carry no strangeness. 
Furthermore, due to its too small number of free parameters,
it artificially  relates the $s/\bar{s}$ asymmetry
 at $x<0.5$ (where NuTeV have significant data) to the one at $x>0.5$
 (where
 NuTeV have few events).
 
 \item  NuTeV did not make a global fit allowing a strange asymmetry,
 but rather made a fit to their dimuon data,  using  a  set of parton
 distributions based on pre-existing (now obsolete) fits,
obtained neglecting NLO QCD corrections and
 optimized under the assumption $s=\bar{s}$. This is especially
 worrisome due to the fact that the strange asymmetry
 found by NuTeV  appears to depend very sensitively on the underlying
 set of parton distributions (see table I of~\cite{Goncharov}).

 \item The dimuon data are considerably less sensitive to $\bar s$ than
 to $s$.  In fact, the claim~\cite{Goncharov,new} that a strange 
asymmetry
 at $x>0.5$ is excluded at high confidence level should be restated as
 the
 statement that NuTeV rules out a {\it total} strangeness at $x>0.5$
of
 the magnitude found by BPZ~\cite{BPZ}.
 However, what matters for the NuTeV anomaly is the strange {\em asymmetry}.

 \end{enumerate}
 
 The BPZ global fit~\cite{BPZ}
 is not subject to these drawbacks, but it did not include
 the recent  dimuon data. The BPZ fit is
  characterized by a relatively large strange sea at
 large  $x$, driven mainly by CDHSW data, which agrees well with
 positivity
 constraints derived from polarized DIS \cite{Forte:2001ph}. Hence, the
 only
 conclusion that can be drawn by comparing these two analyses is that
 the size of the large-$x$ strange sea suggested
 by CDHSW as analyzed by BPZ seems larger than  allowed by
 NuTeV data~\cite{Goncharov}. The origin of this
 discrepancy, and its
 impact on the best-fit
 strange asymmetry, could only be assessed by performing
 a global NLO fit
 which includes all available data. 
 Our  statement in section~3 remains therefore unchanged: 
the impact of
 the data~\cite{Goncharov} on the strange asymmetry is unclear.

 \medskip
 
  NuTeV also comment on isospin violating parton distributions~\cite{new},
 taking the model by Thomas et al.~\cite{Thomas} as reference. 
This model predicts a small
 effect on $\sW^2$, as a result of a subtle cancellation between high
and low $x$ regions.
This conclusion is model-dependent.
The fact remains that ${\cal O}(1\%)$ isospin
 violation
 effects could generate the NuTeV anomaly,
 without conflicting with any other existing data.

 Coming to the possible dilution of strangeness asymmetry or isospin
 violation due to experimental cuts, we should
 like to point out that,
 contrary to what stated in~\cite{new}, we did
 include charm threshold effects and some experimental cuts
 in our analysis. We found moderate
 dilution effects,  as discussed in section 3 of this paper.
 On the other hand, we cannot simulate
 the full NuTeV experimental set-up. However, if
 what NuTeV really measures differs from $R_{\rm PW}$
 in a way which is significant at the required level of accuracy, then
 the cancellation of NLO
 effects that occurs in the Paschos-Wolfenstein relation
 cannot be taken for granted any longer. In particular, NuTeV claims
 to be 
less sensitive to $R_{\bar{\nu}}$ than is $R_{\rm PW}$. 
In general any asymmetry between
 charged-current and neutral-current, 
or between $\nu$ and $\bar \nu$ events  spoils
 the cancellation of the NLO corrections
 in Eq.~(\ref{eq:PWNLO}). Such asymmetries can also be induced by
experimental cuts and   by different
 $\nu$, $\bar{\nu}$ spectra.
If any of these effects were significant, 
a NLO analysis would
 be required in order to obtain a reliable determination of $\sW^2$
 at the desired level of accuracy. Only a full NLO
 analysis of the NuTeV data could settle this issue.

\frenchspacing
\footnotesize
\begin{multicols}{2}

\end{multicols}
\end{document}